\documentclass[%
preprint,
nofootinbib,
 amsmath,amssymb,
 aps,
prd,
longbibliography,
]{revtex4-2}

\usepackage{hyperref}

\usepackage{graphicx}
\usepackage{graphics}
\usepackage{adjustbox} 
\usepackage{dcolumn}
\usepackage{bm}
\usepackage{dsfont}
\usepackage{amsthm}
\usepackage[dvipsnames]{xcolor}
\usepackage{chngcntr}
\usepackage{comment}
\usepackage{apptools}
\AtAppendix{\counterwithin{lemma}{section}}
\usepackage{xy}
\usepackage{qcircuit}
\usepackage{tikz}
\usetikzlibrary{decorations.pathmorphing,decorations.markings}
\usepackage{tkz-euclide}
\usepackage{mathtools}
\usepackage{floatrow} 
\DeclareFloatFont{tiny}{\tiny}
\usepackage{xcolor}

\usepackage[capitalise]{cleveref}
\usepackage[T1]{fontenc}
\crefname{figure}{Fig.}{Fig.}

\theoremstyle{definition}

\theoremstyle{definition}

\def\ket#1{\left\vert #1 \right\rangle}

\newcommand{\be}{\begin{equation}}
\newcommand{\ee}{\end{equation}}
\newcommand{\bp}{\begin{pmatrix}}
\newcommand{\ep}{\end{pmatrix}}
\newcommand{\ben}{\begin{enumerate}}
\newcommand{\een}{\end{enumerate}}

\makeatletter
\newcommand{\beq}{%
 \begingroup
 \eqnarray%
 \@ifstar{\nonumber}{}%
}
\newcommand{\eeq}{\endeqnarray\endgroup}
\makeatother

\newcommand{\pd}{\partial}
\newcommand{\es}{& = &}
\newcommand{\rs}{\ = \ }
\newcommand{\nn}{\nonumber\\}
\newcommand{\np}{\nn & + &}

\newcommand{\nt}{\nn & \times &}

\newcommand{\cA}{\mathcal{A}}
\newcommand{\cB}{\mathcal{B}}
\newcommand{\cC}{\mathcal{C}}
\newcommand{\cD}{\mathcal{D}}
\newcommand{\cF}{\mathcal{F}}
\newcommand{\cG}{\mathcal{G}}
\newcommand{\cH}{\mathcal{H}}

\newcommand{\cM}{\mathcal{M}}
\newcommand{\cN}{\mathcal{N}}
\newcommand{\cP}{\mathcal{P}}
\newcommand{\tdelta}{\tilde\delta}

\usepackage{simpler-wick}
\usepackage{slashed}
\usepackage{ulem}

\begin{document}

\title{Second-order effective renormalized Hamiltonian of Quantum Chromodynamics}

\author{Kamil Serafin}
\affiliation{Department of Physics and Astronomy, Tufts University, Medford, Massachusetts 02155, USA
}%
\author{Carter M. Gustin}
\affiliation{Department of Physics and Astronomy, Tufts University, Medford, Massachusetts 02155, USA
}%
\author{Peter J. Love}
 \altaffiliation[Also at ]{Brookhaven National Laboratory}
\affiliation{Department of Physics and Astronomy, Tufts University, Medford, Massachusetts 02155, USA
}%

\begin{abstract}
The effective Hamiltonian of quantum chromodynamics in the front form of Hamiltonian dynamics is calculated and renormalized. The renormalization group procedure for effective particles up to the second order in the coupling constant is used. Small gluon mass is used to regulate infrared singularities of the theory. The counterterms necessary to renormalize the theory are determined by computing matrix elements of the effective Hamiltonian. The effective Hamiltonians are well-defined symmetric forms on a dense subspace of the Fock space. The zero modes are cut off but, once ultraviolet renormalization is performed, no divergences are found in the color singlet subspace in the limit of the gluon mass approaching zero. A major result is that the interplay between self-energy terms and gluon exchange effective terms generates a term proportional to the quadratic SU(3) Casimir operator times the logarithm of the gluon mass. Therefore, the matrix elements are logarithmically divergent in the color nonsinglet subspace, but finite in the color singlet subspace, because the Casimir operator vanishes in the color singlet subspace. The effective Hamiltonians are suitable for nonperturbative numerical calculations using either classical or quantum computers.
\end{abstract}

\maketitle

%
\section{\label{intro}Introduction}
%

In this paper we compute the effective renormalized
Hamiltonian of QCD up to the second order in the expansion
in the renormalized coupling constant $g$. This Hamiltonian
represents the minimal starting point for a program of
nonperturbative studies of QCD. The main ingredients of
the program include the front form (FF) of Hamiltonian
dynamics~\cite{Brodsky:1997de,Bakker:2013cea},
and the renormalization group procedure for effective
particles (RGPEP)~\cite{Glazek:2000gm,Glazek:2012qj}.

The standard framework for nonperturbative QCD calculations
is lattice QCD, in which field correlation functions are
estimated on lattices of points in Euclidean space using
Monte Carlo sampling methods. Lattice QCD is most successful
with computations of static properties of QCD bound states,
such as masses of hadrons, their magnetic moments, and decay
constants. Scattering properties, such as structure functions
in deep inelastic scattering, are difficult to obtain.
Other calculations face fundamental difficulties due to
the sign problem, which prohibits sampling methods.

In the FF of Hamiltonian dynamics the state of the field
is determined on a three-dimensional hyperplane tangent
to a light cone in Minkowski spacetime. One of the
characteristic properties of the FF is that the center
of mass motion of a system of particles decouples from
the relative motion of these particles. This means that
the relative-motion wave functions of, e.g., quarks in
a proton do not depend on the momentum of the proton as
a whole, which makes our program particularly well-suited
for relativistic scattering calculations. Paired with
the FF is the choice of the light cone gauge which
decouples ghosts and one deals only with physical
degrees of freedom.

The RGPEP is a Wilsonian type of renormalization group
in which one defines a family of effective theories
expressed in terms of effective particles that interact
nonlocally. The family is parameterized by a parameter
$t \ge 0$ and it interpolates between bare theory
with point-like particles interacting locally at $t = 0$
and the diagonal Hamiltonian at $t = \infty$. In this
sense, increasing $t$ brings the theory closer to being
solved.

An additional motivation for developing the QCD program
described in this article is provided by the emerging
quantum computing technologies. Quantum simulation
calculations are done most straightforwardly using
Hamiltonian approaches, hence, FF Hamiltonians can be
applied directly. Quantum computers are expected
to provide new opportunities for the problems that
are most difficult for lattice QCD to solve, including
structure problems of bound states, real-time simulation
of dynamics, and simulation of strongly correlated
matter at high density or far from equilibrium.
The triviality of the FF vacuum also distinguishes
our approach from other proposals such as
Ref.~\cite{Jordan:2012xnu,Jordan:2011ci,Jordan:2014tma}.

In perturbation theory our approach is in principle
equivalent to the standard approaches. The RGPEP
Hamiltonian is obtained from the bare FF Hamiltonian
by means of a unitary transformation. Thus, calculations
for the same quantity will yield the same result as
long as the unitary transformation is calculated to
the same order to which the quantity of interest is
calculated. This independence of parameter $t$ has been
verified explicitly in a relativistic model based on
Yukawa theory for scattering T matrix up to the fourth
order in perturbation
theory~\cite{Maslowski:1997mn,Serafin:2015masters}.
Equivalence of the S matrix for effective and bare
particles has been studied in scalar field theory,
see Ref.~\cite{Wieckowski:2005gp}.

The FF coordinates cannot be related to any inertial
frame via a Lorentz transformation, hence, the FF
is a substantially distinct framework. Nevertheless
one can compute correlation functions of fields using
perturbation theory equally well in both the standard
instant form and the FF using various techniques, such
as Feynman diagrams, or old-fashioned perturbation
theory. Although the final results are the same,
the intermediate steps are done differently, and special
attention needs to be given to the so-called zero mode
contributions, whose proper inclusion can be tricky~\cite{Collins:2018aqt,Chang:1968bh,Mannheim:2020rod,Polyzou:2021qpr}.
The physical picture of certain phenomena can also differ
substantially between different forms of dynamics. The most
significant difference in QCD is the role that the vacuum
plays in the appearance of confinement and chiral symmetry
breaking. In the instant form, both confinement and chiral
symmetry breaking are directly linked to the complicated
structure of the vacuum via quark and gluon condensates.
In the front form all complexities of the vacuum state have
to be contained in the special subset of field modes that
have exactly zero longitudinal momentum---these excitations
are called zero modes. Furthermore, one can remove the zero
modes from the theory, but restore all of their effects
on the theory by including appropriate counterterms in the
Hamiltonian (and possibly other operators of interest).
Therefore, a complicated vacuum in the instant form is
replaced with a trivial vacuum and special counterterms
in the Hamiltonian in the front form. In the field theory
of a scalar field the effects of zero modes amount to
unobservable shifts of bare masses and coupling constants.
To the best of our knowledge, the zero-mode counterterms
in QCD, whose inclusion is necessitated by the removal
of the zero modes, are not fully known. One way of studying
these problems is the discretized light-cone quantization
(DLCQ), in which the theory is quantized in a box.
Numerical solutions for 1+1D QCD done using DLCQ provide
in fact a complete solution of the theory~\cite{Hornbostel:1988ne,Hornbostel:1988fb}. Analogous
calculations in 1+3-dimensional QCD are much more demanding
in terms of computational resources needed to produce
high-quality results. However, if given enough resources,
the approach might be equally successful in 1+3D as it was
in 1+1D. The multitude of results, and the
continued progress done by the BLFQ Collaboration
supports this claim by example~\cite{Xu:2024sjt}. While the main goal of
the BLFQ Collaboration at the moment is the study of the
bare canonical Hamiltonian of QCD, we provide an effective,
i.e., renormalized Hamiltonian that can be studied using
similar methods. Our Hamiltonian is a minimal, well-defined
starting point for nonperturbative calculations in QCD.
To study the influence of the missing zero-mode counterterms
one might perhaps need to use methods similar to those
described in Ref.~\cite{Chabysheva:2025vtm}. In any case,
numerical calculations are supposed to guide the development
of the theory of the front-form zero-mode counterterms.
Therefore, we make calculations in a way that should aid
subsequent studies of all the important properties of
front-form Hamiltonians, including confinement, chiral
symmetry breaking, and renormalizability.

References~\cite{Tomboulis:1973jn,Casher:1976ae} represent an early work on FF QCD and important applications to exclusive processes were investigated in Ref.~\cite{Lepage:1980fj}. Various problems of the canonical FF QCD Hamiltonian have been studied in Refs.~\cite{Zhang:1993is,Zhang:1993dd,Harindranath:1993de}, including coupling constant renormalization and asymptotic freedom~\cite{Harindranath:1993de,Perry:1992sw}. Recent numerical studies within the framework of basis light-front quantization (BLFQ)~\cite{Zhao:2020kuf,Vary:2025yqo} use Fock-sector-dependent renormalization~\cite{Perry:1990mz,Perry:1991ny,Karmanov:2008br}. Other authors emphasize regulators that preserve Lorentz and gauge symmetries at the expense of introducing Pauli-Villars ghost fields~\cite{Hiller:2016itl,Paston:1999dst}. Chiral symmetry and constrained fermion zero modes in QCD that induce effective interactions were studied in, e.g., Ref.~\cite{Dalley:2004re}.

Investigations of the QCD Hamiltonian using the similarity renormalization group (SRG)~\cite{Glazek:1993rc,Glazek:1994qc} begin with Ref.~\cite{Wilson:1994fk}. Important findings following the approach of~\cite{Wilson:1994fk} include cancellation of infrared divergences and the presence of logarithmic confining potentials within the color-singlet Fock subspace~\cite{Perry:1994mv}. These potentials rise twice as quickly in the transverse direction as they do in the longitudinal direction. In contrast, in 1+2-dimensional QCD the analogous potentials are linear in the transverse direction, while of the square root form in the longitudinal direction~\cite{Chakrabarti:2001iq}. Bound state studies of heavy-light mesons and heavy quarkonia using early versions of SRG have been pursued~\cite{Brisudova:1995hv,Brisudova:1996vw}. Following the principle of coupling coherence~\cite{Perry:1993gp,Perry:2001je}, Yang-Mills Hamiltonian, and spectra and wave functions of glueballs have been found~\cite{Allen:1999kx}. Within heavy-flavor QCD an approach based on the gluon mass ansatz has been proposed~\cite{Glazek:2003ky}. In principle, infinitely many Fock sectors can be integrated out. The ansatz is that gluons in the remaining sectors acquire an effective mass which crudely represents effects of nonperturbative partial diagonalization necessary to reduce the Hamiltonian. This approach has been applied to heavy quarkonia~\cite{Glazek:2006cu,Glazek:2017rwe}, triply heavy baryons~\cite{Serafin:2018aih}, as well as glueballs~\cite{Maslowski:2005phd}. For separations between heavy quarks typical for the ground states of heavy quarkonia, the confining potentials are approximately harmonic and rotationally symmetric as opposed to the long-distance logarithmic behavior~\cite{Serafin:2023pkf}. Using a form of the RGPEP closely resembling our calculations, asymptotic freedom has been confirmed in a calculation of the effective triple-gluon vertex~\cite{Gomez-Rocha:2015esa}.

In a previous article, Ref.~\cite{Serafin:2025ouo}, we computed the second-order effective Hamiltonian of Yukawa theory and we introduced a technique based on Wick's diagrams in momentum space. Wick's diagrams are meant to formalize a graphical representation of the mathematical formulas for interaction operators in the effective Hamiltonian. Importantly, a single diagram typically represents several distinct time orderings, e.g., gluon splitting into two gluons and two gluons merging into a single gluon are represented with one triple-gluon Wick's diagram. This method has been developed to the extent necessary at the second order of perturbation theory for the effective Hamiltonians.

In this article we use methods developed in Ref.~\cite{Serafin:2025ouo} to compute the full expression for the second-order effective Hamiltonian of QCD on the light-front, thereby extending the results of Ref.~\cite{Serafin:2023pkf} to all Fock sectors. We start with the canonical FF Hamiltonian of QCD and remove zero modes from interactions with longitudinal momentum cutoff $p^+ > \epsilon^+$. This ensures triviality of the vacuum but requires inclusion of zero-mode counterterms that will fully reproduce effects of confinement and chiral symmetry breaking---a task that we postpone to a later work. The zero-mode counterterms may shift bare masses and coupling constants. Therefore, at least part of the zero-mode counterterms can be recovered by simply fitting the unperturbed masses, and the coupling constant to reproduce a small set of observables (a step that is reasonable in any case). Analogous shifts in the scalar theory, in fact, include all effects of the removed zero modes~\cite{Burkardt:1992sz,Burkardt:2000jn}. In the same vein, since chiral symmetry breaking leads to quarks acquiring large effective masses, allowing large values for quark masses in the fitting process will make it possible to reproduce at least some effects of chiral symmetry breaking.

Following Ref.~\cite{Serafin:2023pkf}, we also introduce a gluon mass, $m_g$ to regulate the singularities that arise when longitudinal momentum of a gluon approaches zero. This regulator likely breaks gauge invariance of correlation functions computed perturbatively using the effective Hamiltonians. Therefore, having renormalized the ultraviolet divergences in the Hamiltonian, we perform the $m_g \to 0$ limit, which will restore the standard results (to the extent that the canonical Hamiltonian without zero-mode counterterms can produce). Gauge symmetry is not manifest in our approach from the very start, since we fix the gauge to the light-cone gauge. The main purpose of the effective Hamiltonians that we compute is to use them in nonperturbative calculations using, e.g., DLCQ or BLFQ. For that purpose, breaking of gauge and Lorentz symmetry is acceptable~\cite{Wilson:1994fk}. In fact, the choice of the RGPEP generator that we make further breaks the exact longitudinal boost invariance of the FF. This is due to the approximate  nature of the solutions of the RGPEP equations that we compute. If we could solve these equations exactly, the longitudinal boost invariance would also be exact.

Despite the omission of the zero-mode counterterms removal of the infrared regularization in the renormalized effective Hamiltonian, i.e., the $m_g \to 0$ limit, yields a well-defined theory in the color-singlet Fock subspace. This is because all terms that depend on the gluon mass are proportional to a Casimir operator of $SU(3)$, which vanishes in the singlet sector. This is in apparent contradiction with Refs.~\cite{Zhang:1993is,Wilson:1994fk}. Additionally, as opposed to Ref.~\cite{Wilson:1994fk}, we do not include any universally confining potential, whose inclusion in Ref.~\cite{Wilson:1994fk} was motivated by a potential instability of the Hamiltonian. At this point there appears no need for such measures. Numerical simulations should detect the instability, should it be present in our Hamiltonians.

The paper is organized as follows. The canonical Hamiltonian of QCD is reviewed and notation is established in Sec.~\ref{sec:canonical}. In Sec.~\ref{sec:regularized} we discuss the regulating factors that we introduce. The effective Hamiltonian is computed up to the second order in Sec.~\ref{sec:rgpep}. In Sec.~\ref{sec:counterterms} we study the matrix elements of the effective Hamiltonian and determine the counterterms that need to be added to the bare Hamiltonian to make the effective Hamiltonian well defined in the limit of removing the ultraviolet regularization. Section~\ref{sec:casimir} concerns the limit of removing the infrared regulating gluon mass and the quadratic Casimir operator is found. Section~\ref{sec:summary} summarizes the effective Hamiltonian, and we make concluding remarks in Sec.~\ref{sec:conclusion}.

%
\section{Canonical Hamiltonian of QCD}
\label{sec:canonical}
%

The canonical Hamiltonian is the integral of the density over
the hypersurface defined by $x^+ = 0$,
\beq
H_\text{canonical}
\es
\int dx^- d^2 x^\perp \, \cH
\ .
\eeq
The density of the canonical Hamiltonian of QCD
in the Front Form of Hamiltonian dynamics is,
\beq
\cH
\es
  \cH_{\psi^2 + A^2} + \cH_{jA} + \cH_{A^4}
+ \cH_{\psi A A \psi} + \cH_{jj}
\ ,
\eeq
where
\beq
\cH_{\psi^2 + A^2}
\es
  \cN\left( \bar\psi \frac{\gamma^+}{2}
  \frac{ (i\pd^\perp)^2 + m^2}
  {i\pd^+}\psi
  \right)
+ \cN\left\{
  \frac{1}{2} A^{i\,a}
  \left[ m_g^2 + (i\pd^\perp)^2 \right] A^{i\,a}
  \right\}
 ,
\\
\cH_{jA}
\es
  j_{q\,\mu}^a A^{\mu a}
+ j_{g\,\mu}^a A^{\mu a}
\ ,
\\
\cH_{A^4}
\es
  \frac{1}{4} g^2 f^{abc} f^{ade}
  \cN\left( A_\alpha^b A_\beta^c
  A^{\alpha\,d} A^{\beta\,e} \right)
 ,
\label{eq:denHA4}
\\
\cH_{\psi A A \psi}
\es
\frac{1}{2} g^2
\cN\left(
\bar\psi \gamma^i A^{i a} T^a
\frac{\gamma^+}{i\pd^+} \gamma^j A^{j b} T^b \psi
\right)
 ,
\\
\cH_{jj}
\es
\frac{1}{2}
\cN\left[
\left( j_q^{+a} + 3 j_g^{+a} \right)
\frac{1}{(i\pd^+)^2}
\left( j_q^{+a} + 3 j_g^{+a} \right)
\right]
 ,
\label{eq:cHjj}
\eeq
$\psi$ is the quark field and $A$ is the gluon field, and
$j_q^{\mu a}$ and $j_g^{\mu a}$ are hermitian color current
operators,
\beq
j_q^{\mu a}
\es
g
\cN\left( \bar\psi \gamma^\mu T^a \psi \right) ,
\\
j_g^{\mu c}
\es
g^{\mu\gamma}
\frac{g}{3!}
i f^{abc}
\left\{
  g_{\alpha\beta}
  \left[
    A^{\alpha a}(x) i \pd_\gamma A^{\beta b}(x)
  - i \pd_\gamma A^{\alpha a}(x) A^{\beta b}(x)
  \right]
\right.
\nn&&
- g_{\beta\gamma}
  \left[
    i \pd_\alpha A^{\alpha a}(x) A^{\beta b}(x)
  + 2 A^{\alpha a}(x) i \pd_\alpha A^{\beta b}(x)
  \right]
\nn&&
\left.
+ g_{\gamma\alpha}
  \left[
    2 i \pd_\beta A^{\alpha a}(x) A^{\beta b}(x)
  + A^{\alpha a}(x) i \pd_\beta A^{\beta b}(x)
  \right]
\right\} .
\label{eq:jgmua}
\eeq
Indices $a, b, c, d, e = 1, \dots, 8$ denote color.
Indices $i, j = 1, 2$ denote the components of
the transverse momentum two-vectors.

The definition, Eq.~(\ref{eq:jgmua}) differs from
other, notable definitions used before. For example,
for the purpose of defining the canonical Hamiltonian
one could use a nonsymmetric gluon color current
of the form $j_\text{NS}^{\mu\,a} = - i g f^{abc} A^{jb}
i\pd^\mu A^{jc}$, which is the same as the gluon
color current $\tilde\chi^{\mu\,a}$ defined
in Ref.~\cite{Brodsky:1997de}, see Eq.~(2.79) there.
The Hamiltonian density, $\cH$, as defined above
with the use of $j_g^{\mu\,a}$, is the same as
that given in Ref.~\cite{Brodsky:1997de}, where
$\tilde\chi^{\mu\,a}$ was used. Since,
$j_\text{NS}^{+\,a} = 3 j_{g}^{+\,a}$, using
$j_\text{NS}^{\mu\,a}$ is convenient when writing down
instantaneous Hamiltonian term $\cH_{jj}$, because there
would be no need for the factors of 3 in Eq.~(\ref{eq:cHjj}).
Our choice leads more directly to the symmetric
form of the three-gluon vertex in momentum space,
see Eq.~(\ref{eq:3gluonCan}). This choice also fits
better the structure of later calculations, in which
two three-gluon vertices are contracted with each other,
and a symmetry factor needs to be included because
contractions of different legs of the two vertices
leads to the same Hamiltonian terms. In this sense,
the canonical instantaneous-gluon interaction $\cH_{jj}$
is interpreted as composed of two three-leg diagrams.
If either of the diagrams is a three-gluon vertex,
then there is a three-fold choice of which of the gluon
legs is to be contracted.\footnote{$j_{g\,\mu}^a A^{\mu a}$
is not a contraction of two diagrams, and does not require
any symmetry factors.} Therefore, calculations in which
symmetric three-gluon vertices are used, and in which
the instantaneous vertices can be interpreted as
contractions of two three-leg interactions, seem more
consistent. Although in the second-order calculations
we already see the utility of our choice, we anticipate
that the effort of keeping track of all the factors
of 3 will be most useful in calculations of order
higher than the second.

The quark field is assigned mass $m$, while the gluon
field is assigned mass $m_g$. The gluon mass is an
important modification of the usual QCD Hamiltonian
introduced to regulate singularities associated with
the vanishing longitudinal momentum of gluons.
The fields are expanded in the plane wave basis,
\beq
\psi(x)
\es
\sum_{c, \sigma} \int\frac{dp^+ d^2p^\perp}{16\pi^3 p^+} \theta(p^+)
\left[
  \chi_c u_\sigma(p) e^{-i p x} b_{p \sigma c}
+ \chi_c v_\sigma(p) e^{ i p x} d_{p \sigma c}^\dag
\right]
 ,
\\
A^{\mu\,c}(x)
\es
\sum_\sigma \int\frac{dp^+ d^2p^\perp}{16\pi^3 p^+} \theta(p^+)
\left[
  \varepsilon_\sigma^\mu(p) e^{-i p x} a_{p \sigma c}
+ \varepsilon_\sigma^{\mu*}(p) e^{i p x} a_{p \sigma c}^\dag
\right]
 ,
\eeq
where $b_{p \sigma c}$, and $d_{p \sigma c}$ are annihilation
operators for a quark and an antiquark with momentum $p$,
light-front helicity, i.e., spin projection onto $z$ axis,
$\sigma = \pm \frac{1}{2}$, and color $c = 1, 2, 3$,
respectively. Gluon with momentum $p$, light-front helicity
$\sigma = \pm 1$, and color $c = 1, \dots, 8$ is annihilated
with an operator $a_{p \sigma c}$. With quarks and antiquarks
are associated spinors $u_\sigma(p)$ and $v_\sigma(p)$, which
are solutions of the free Dirac equation, and the color
three-vector $\chi_c = [ \delta_{c,1}, \delta_{c,2},
\delta_{c,3} ]^T$, where $T$ denotes transposition. Gluons
have a polarization four-vector $\varepsilon^\mu_\sigma(p)$
associated with them, and the star in $\varepsilon^{\mu*}_\sigma(p)$
means complex conjugation. $\varepsilon_\sigma^+(p) = 0$ and
$p_\mu\varepsilon_\sigma^\mu(p) = 0$. Since the
longitudinal component of the angular momentum is
conserved exactly in the front form, circular
polarization vectors, e.g., $\varepsilon^\perp_{\pm 1}
= [\mp 1, -i]^T/\sqrt{2}$, are a useful choice.

We define the Fourier transforms, $\Psi(q)$, $G^{\mu\,a}(q)$,
$J_q^{\mu\,a}(q)$, and $J_g^{\mu\,a}(q)$ of $\psi(x)$,
$A^{\mu\,a}(x)$, $j_q^{\mu\,a}(q)$, and $j_g^{\mu\,a}(q)$,
respectively, in accordance with the generic formula,
\beq
F(q)
\es
\int dx^- d^2 x^\perp
e^{\frac{i}{2} q^+ x^- - i q^\perp x^\perp}
f(x)
\ .
\eeq
Thus,
\beq
\Psi(q)
\es
\sum_{c, \sigma}
\frac{
  \theta(q^+) \chi_c u_{\sigma}(q) b_{q \sigma c}
+ \theta(-q^+) \chi_c v_{\sigma}(-q) d_{-q \sigma c}^\dag
}{|q^+|}
\ ,
\\
G^{\mu \, a}(q)
\es
\sum_\sigma
\frac{
\theta(q^+)
\varepsilon_\sigma^\mu(q)
a_{q \sigma c}
+ \theta(-q^+)
\varepsilon_\sigma^{\mu*}(-q)
a_{-q \sigma c}^\dag
}{|q^+|}
\ ,
\\
J_{q/g}^{\mu \, a}(q)
\es
\int[q_1 q_2]
\,\tdelta_{12.q}
\,\tilde J_{q/g}^{\mu \, a}(q_1, q_2)
\ ,
\eeq
where
\beq
\tilde J_q^{\mu \, a}(q_1, q_2)
\es
g
\,\cN\left[
\bar\Psi(-q_1) \, \gamma^\mu \, T^a \, \Psi(q_2)
\right] ,
\label{eq:Jq}
\\
\tilde J_g^{\mu \, c}(q_1, q_2)
\es
\, g^{\mu\gamma}
\frac{g}{3!}
\, F_{\alpha\beta\gamma}^{abc}(q_1, q_2, - q_1 - q_2)
\, G^{\alpha \, a}(q_1) \, G^{\beta \, b}(q_2) \ ,
\label{eq:Jg}
\eeq
and
\beq
F_{\alpha\beta\gamma}^{abc}(q_1, q_2, q_3)
\es
i f^{abc}
\left[
  (q_{2\gamma} - q_{1\gamma}) g_{\alpha\beta}
+ (q_{3\alpha} - q_{2\alpha}) g_{\beta\gamma}
+ (q_{1\beta} - q_{3\beta}) g_{\gamma\alpha}
\right] .
\eeq
$F_{\alpha\beta\gamma}^{abc}(q_1, q_2, q_3)$ differs from
the usual three-gluon vertex factor used in Feynman diagrams
only by a constant factor. The current densities satisfy
\beq
\tilde J_{q/g}^{\mu\,c}(q_1, q_2)^\dag
\es
\tilde J_{q/g}^{\mu\,c}(-q_2, -q_1)
\ ,
\\
\tilde J_{g}^{\mu\,c}(q_2, q_1)
\es
\tilde J_{g}^{\mu\,c}(q_1, q_2)
\ ,
\\
(q_{1\mu} + q_{2\mu}) \tilde J_{q/g}^{\mu\,c}(q_1, q_2)
\es
0
\ ,
\\
q_\mu \tilde J_{q/g}^{\mu\,c}(q_1, q_2)
\es
\frac{q^- - q_1^- - q_2^-}{2}
\tilde J_{q/g}^{+\,c}(q_1, q_2)
\ ,
\label{eq:qmuJmu}
\eeq
with $q^{+,\perp} = q_1^{+,\perp} + q_2^{+,\perp}$, but
$q^- \neq q_1^- + q_2^-$ in general, i.e., $q^\mu =
q_1^\mu + q_2^\mu + n^\mu (q^- - q_1^- - q_2^-)/2$,
where $n^\mu$ is a null fourvector tangent to the light
front hypersurface such that $n^- = 2$, and $n^+ = n^1 =
n^2 = 0$, while the nonzero elements of the metric tensor
are $g^{+-} = g^{-+} = 2$, $g^{11} = g^{22} = -1$.
To prove Eq.~(\ref{eq:qmuJmu}) for quark current density
Dirac equation, $\slashed{q} \Psi(q) = m \Psi(q)$ is
used, while for gluon current density $q_\mu G^{\mu\,c}(q)
= 0$, and $q_{1\mu} q_1^\mu = q_{2\mu} q_2^\mu$ are used.

Note that $q^+$ can be both
positive and negative and the Fourier transform is
three-dimensional, hence, $\Psi$ and $G$ are functions
of $q^+$ and $q^\perp$ only. One can, however, assign
$q^-$ components in accordance to free evolution,
\beq
i\pd_f^- \Psi(q)
\es
\left.q^-\right|_{m} \Psi(q)
\ ,
\\
i\pd_f^- \bar\Psi(-q)
\es
\left.q^-\right|_{m} \bar\Psi(-q)
\ ,
\\
i\pd_f^- G(q)
\es
\left.q^-\right|_{m_g} G(q)
\ ,
\eeq
where $\left.q^-\right|_m = \frac{m^2 + (q^\perp)^2}{q^+}$,
$\left.q^-\right|_{m_g} = \frac{m_g^2 + (q^\perp)^2}{q^+}$.

The canonical Hamiltonian is,
\beq
H_\text{canonical}
\es
  H_{\psi^2 + G^2} + H_{\psi^2 G}
+ H_{G^3} + \cH_{G^4}
+ H_{\psi^2 G^2} + H_{JJ}
\ .
\eeq
It consists of the free part, and first- and second-order
vertices. For illustration purposes we represent interaction
terms as Wick's diagrams~\cite{Serafin:2025ouo}.
We also introduce simplified notation:
\beq
\tdelta_{1\dots n}
\rs
16\pi^3 \delta(q_1^+ + \dots + q_n^+)
\delta^2(q_1^\perp + \dots + q_n^\perp)
\ ,
\\
\left[q\right] \rs \frac{dq^+ d^2 q^\perp}{16\pi^3} \ ,
\quad
[q_1 q_2] \rs [q_1] [q_2] \ ,
\quad\text{etc.}
\eeq

The free part  of the Hamiltonian is,
\beq
H_{\psi^2 + G^2}
\es
  \int[q]\,
  \left.q^-\right|_m
  \cN\left[
  \bar\Psi(q)
  \frac{\gamma^+}{2}
  \Psi(q)
  \right]
+
  \int[q]\,
  q^+ \left.q^-\right|_{m_g}
  \cN\left[
  \frac{1}{2} G^{i\,a}(-q) G^{i\,a}(q)
  \right] .
\label{eq:freeCan}
\eeq

\begin{figure}
 \centering
 \includegraphics{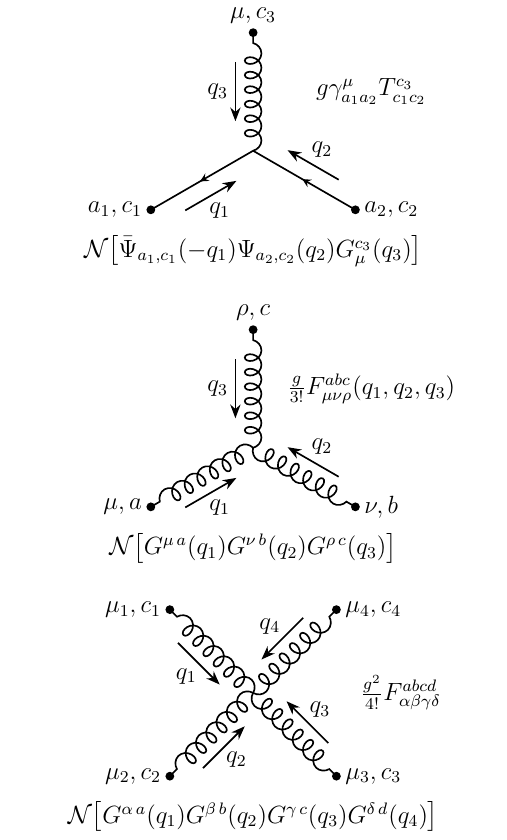}
 \caption{Basic diagrams.}
 \label{fig:basic}
\end{figure}

There are two first-order vertices. The quark-gluon vertex is,
\beq
H_{\psi^2 G}
\es
g
\int[q_1 q_2 q_3]
\,\tdelta_{123}
\,\cN\left[
\bar\Psi(-q_1) \gamma_\mu T^a \Psi(q_2) G^{\mu a}(q_3)
\right] ,
\eeq
and it is represented in Fig.~\ref{fig:basic}(a).
The triple gluon vertex is,
\beq
H_{G^3}
\es
\frac{g}{3!}
\int[q_1 q_2 q_3]
\,\tdelta_{123}
F_{\alpha\beta\gamma}^{abc}(q_1, q_2, q_3)
\, G^{\alpha a}(q_1) G^{\beta b}(q_2) G^{\gamma c}(q_3)
\ .
\label{eq:3gluonCan}
\eeq
It is represented as a Wick's diagram in Fig.~\ref{fig:basic}(b).

\begin{figure}
 \centering
 \includegraphics{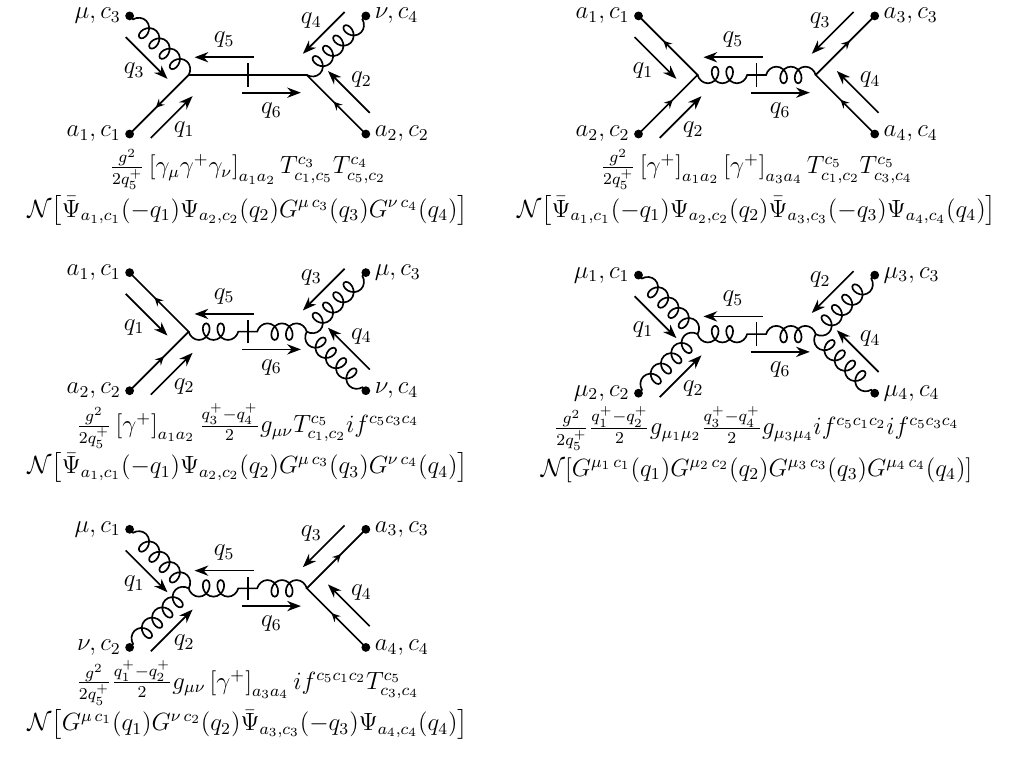}
 \caption{Instantaneous diagrams.}
 \label{fig:instantaneous}
\end{figure}

There are three second-order vertices. The first one is
the usual four-gluon vertex,
\beq
H_{G^4}
\es
\frac{g^2}{4!}
\int[q_1 q_2 q_3 q_4]
\,\tdelta_{1234}
\,F_{\alpha\beta\gamma\delta}^{abcd}
\,\cN\left[
G^{\alpha\,a}(q_1)
G^{\beta\,b}(q_2)
G^{\gamma\,c}(q_3)
G^{\delta\,d}(q_4)
\right] ,
\eeq
where
\beq
F_{\alpha\beta\gamma\delta}^{abcd}
\es
  f^{abe} f^{cde}
  ( g_{\alpha\gamma} g_{\beta\delta}
  - g_{\alpha\delta} g_{\beta\gamma} )
+ f^{ace} f^{bde}
  ( g_{\alpha\beta} g_{\gamma\delta}
  - g_{\alpha\delta} g_{\beta\gamma} )
+ f^{ade} f^{bce}
  ( g_{\alpha\beta} g_{\gamma\delta}
  - g_{\alpha\gamma} g_{\beta\delta} )
\ .
\nn
\eeq
The four-gluon vertex is symmetrized from its nonsymmetric
form of Eq.~(\ref{eq:denHA4}). This interaction term is
represented as a Wick's diagram in Fig.~\ref{fig:basic}(c).
The other two second-order vertices are characteristic to
the front form and are called instantaneous. The instantaneous
fermion vertex is,
\beq
H_{\psi^2 G^2}
\es
\frac{1}{2} g^2
\int[q_1 q_2 q_3 q_4]
\,\tdelta_{1234}
\,\cN\left[
\bar\Psi(-q_1) \gamma^i G^{i a}(q_2) T^a
\frac{\gamma^+}{q_3^+ + q_4^+}
\gamma^j G^{j b}(q_3) T^b \Psi(q_4)
\right] .
\eeq
It is represented as a Wick's diagram in
Fig.~\ref{fig:instantaneous}(a).
The instantaneous gluon interaction is,
\beq
H_{JJ}
\es
\frac{1}{2}
\int[q]\,
\cN\left\{
\left[ J_q^{+a}(-q) + 3 J_g^{+a}(-q) \right]
\frac{1}{(q^+)^2}
\left[ J_q^{+a}(q) + 3 J_g^{+a}(q) \right]
\right\} .
\label{eq:HcanJJ}
\eeq
This vertex has four contributions represented as Wick's
diagrams in Fig.~\ref{fig:instantaneous}(b)--(e).

%
\section{Regularized Hamiltonian of QCD}
\label{sec:regularized}
%

The first cutoff we introduce is the small $p^+$ cutoff
in the Fourier expansion of fields,
\beq
\psi(x)
\es
\sum_{c, \sigma} \int\frac{dp^+ d^2p^\perp}{16\pi^3 p^+} \theta_\epsilon(p^+)
\left[
  \chi_c u_\sigma(p) e^{-i p x} b_{p \sigma c}
+ \chi_c v_\sigma(p) e^{ i p x} d_{p \sigma c}^\dag
\right]
 ,
\\
A^{\mu\,c}(x)
\es
\sum_\sigma \int\frac{dp^+ d^2p^\perp}{16\pi^3 p^+}
\theta_\epsilon(p^+)
\left[
  \varepsilon_\sigma^\mu(p) e^{-i p x} a_{p \sigma c}
+ \varepsilon_\sigma^{\mu*}(p) e^{i p x} a_{p \sigma c}^\dag
\right]
 ,
\eeq
where $\theta_\epsilon(p^+) = \theta(p^+ - \epsilon^+)$, and $\epsilon^+ > 0$ is the minimal $p^+$ any mode can have.
Therefore,
\beq
\Psi(q)
\es
\sum_{c, \sigma}
\frac{
  \theta_\epsilon(q^+) u_{\sigma}(q) b_{q \sigma c}
+ \theta_\epsilon(-q^+) v_{\sigma}(-q) d_{-q \sigma c}^\dag
}{|q^+|}
\chi_c
\ ,
\\
G^{\mu \, c}(q)
\es
\sum_\sigma
\frac{
  \theta_\epsilon(q^+)
  \varepsilon_\sigma^\mu(q)
  a_{q \sigma c}
+ \theta_\epsilon(-q^+)
  \varepsilon_\sigma^{\mu*}(-q)
  a_{-q \sigma c}^\dag
  }{|q^+|}
\ .
\eeq
The purpose of this cutoff is to remove interactions
that would produce zero modes from the vacuum, thus
ensuring the triviality of the vacuum state. 
The limit $\epsilon^+ \to 0^+$ is the first limit
to be taken.

Local interactions in QCD lead to ultraviolet divergences,
which make the canonical Hamiltonian not well defined.
Hence, we redefine the interaction terms of the canonical
Hamiltonian. The three-leg interactions become,
\beq
H_{\psi^2 G}
\es
g
\int[q_1 q_2 q_3]
\, f_{t_r,123}^q
\,\tdelta_{123}
\,\cN\left[
\bar\Psi(-q_1) \gamma_\mu T^a \Psi(q_2) G^{\mu a}(q_3)
\right]
\ ,
\\
H_{G^3}
\es
\frac{g}{3!}
\int[q_1 q_2 q_3]
\, f_{t_r,123}^g
\,\tdelta_{123}
F_{\alpha\beta\gamma}^{abc}(q_1, q_2, q_3)
\, G^{\alpha a}(q_1) G^{\beta b}(q_2) G^{\gamma c}(q_3)
\ ,
\eeq
where $t_r$ is the cutoff parameter, and $f_{t_r,123}^q$
and $f_{t_r,123}^g$ are regulating functions,
\beq
f_{t_r,123}^{q}
\es
\exp\left[
- t_r
  \left(
      \left.q_1^-\right|_{m}
    + \left.q_2^-\right|_{m}
    + \left.q_3^-\right|_{m_g}
  \right)^2
\right] .
\label{eq:ffq}
\eeq
and
\beq
f_{t_r,123}^{g}
\es
\exp\left[
- t_r
  \left(
      \left.q_1^-\right|_{m_g}
    + \left.q_2^-\right|_{m_g}
    + \left.q_3^-\right|_{m_g}
  \right)^2
\right] .
\label{eq:ffg}
\eeq
The subscripts in $f_{t_r,123}^{q}$ and $f_{t_r,123}^{g}$ are not to be confused as
containing number 123, but rather three numbers, 1, 2, and 3
that represent momenta of particles 1, 2, and 3, respectively.
Note that the order of subscripts matters. The difference
between $f_{t_r,123}^q$ and $f_{t_r,123}^g$ is in the definition
of the minus components of $q_i$s. For $f_{t_r,123}^q$,
the definition of $q_1^-$, $q_2^-$, and $q_3^-$ ensures
$q_1^2 = q_2^2 = m^2$, and $q_3^2 = m_g^2$. For $f_{t_r,123}^g$,
the definition of $q_1^-$, $q_2^-$, and $q_3^-$ ensures
$q_1^2 = q_2^2 = q_3^2 = m_g^2$. If one can infer from
the context what the masses that need to be used for each
$q_i^-$ are, then one can drop the mass annotation of $q_i^-$
and simply write $f_{t_r,123}$, without the superscript $q$
or $g$. Note that this may lead to false simplifications,
e.g., $f_{t_r,123}^q A + f_{t_r,123}^g B \neq f_{t_r,123}
(A+B)$. For clarity we will keep the superscripts. The limit $t_r \to 0$ makes $f^{q/g}_{t_r,123} \to 1$,
and it corresponds to the removal of the regularization.

The four-gluon vertex is redefined with a similar
regulating function,
\beq
H_{G^4}
\es
\frac{g^2}{4!}
\int[q_1 q_2 q_3 q_4]
\,f_{t_r,1234}
\,\tdelta_{1234}
\,F_{\alpha\beta\gamma\delta}^{abcd}
\,\cN\left[
G^{\alpha\,a}(q_1)
G^{\beta\,b}(q_2)
G^{\gamma\,c}(q_3)
G^{\delta\,d}(q_4)
\right] ,
\eeq
where
\beq
f_{t_r,1234}
\es
\exp\left[
- t_r
  \left(
      \left.q_1^-\right|_{m_g}
    + \left.q_2^-\right|_{m_g}
    + \left.q_3^-\right|_{m_g}
    + \left.q_4^-\right|_{m_g}
  \right)^2
\right] .
\eeq
The instantaneous fermion vertex is,
\beq
H_{\psi^2 G^2}
\es
\frac{1}{2} g^2
\int[q_1 q_2 q_3 q_4 q_5]
\,\tdelta_{1234}
\,\tdelta_{34.5}
f^q_{t_r,1(-5)2}
f^q_{t_r,(-5)43}
\nn&&\times
\cN\left[
\bar\Psi(-q_1) \gamma^i G^{i a}(q_2) T^a
\frac{\gamma^+}{q_5^+}
\gamma^j G^{j b}(q_3) T^b \Psi(q_4)
\right] ,
\label{eq:HregPsi2G2}
\eeq
where
\beq
\tdelta_{34.5}
\rs
16\pi^3 \delta(q_3^+ + q_4^+ - q_5^+)
\delta^2(q_3^\perp + q_4^\perp - q_5^\perp)
\ ,
\eeq
and $-5$ in $f^q_{t_r,1(-5)2}$ and in $f^q_{t_r,(-5)43}$
means that $q_5$ momentum is to be taken with minus sign
in Eq.~(\ref{eq:ffq}). Momentum $q_5$ is introduced
by adding one more integration and one more Dirac delta,
and it serves two purposes. Firstly, $q_5^+$ is a place
holder for $q_3^+ + q_4^+$. Secondly, $q_5^-$ (which is not
equal to $q_3^- + q_4^-$) in the regulating function
$f^q_{t_r,1(-5)2} f^q_{t_r,(-5)43}$, makes the regularization
act as if there were an actual particle between the two
three-leg vertices depicted in Fig.~\ref{fig:instantaneous}(a).

The instantaneous gluon vertex is redefined to be,
\beq
H_{JJ}
\es
\frac{1}{2}
\int[q]
\,\cN\left\{
\left[ J_{q,t_r}^{+a}(-q) + 3 J_{g,t_r}^{+a}(-q) \right]
\frac{1}{(q^+)^2}
\left[ J_{q,t_r}^{+a}(q) + 3 J_{g,t_r}^{+a}(q) \right]
\right] ,
\label{eq:HregJJ}
\eeq
where
\beq
J_{q/g,t_r}^{\mu \, a}(q)
\es
g
\int[q_1 q_2]
\, \tdelta_{12.q}
\, f^{q/g}_{t_r,12(-q)}
\, \tilde J_{q/g}^{\mu \, a}(q)
\ .
\eeq
$q$ does not have a subscript, hence in $\tdelta_{12.q}$
and $f^{q/g}_{t_r,12(-q)}$ we use $q$ itself instead of
its subscript. These expressions are constructed in exactly
the same way as they would be with a subscripted momentum
(such as $q_5$ in $\tdelta_{34.5}$ and $f^q_{t_r,1(-5)2}$).
The instantaneous gluon vertex is regulated as if it were
composed of two three-leg vertices. This makes it consistent
with the regularization functions encountered in composite
terms obtained by Wick's contractions, which in terms of
Wick's diagrams means connecting individual Wick's diagrams
into composite ones.

Due to the gluon mass, $m_g$ the UV regularization factors,
$f^{q}_{t_r,123}$, $f^{g}_{t_r,123}$ regulate also what may be considered
infrared singularities, i.e., factors of longitudinal
momentum in the denominator, such as $1/q_5^+$ in
Eq.~(\ref{eq:HregPsi2G2}) or $1/(q^+)^2$ in
Eq.~(\ref{eq:HregJJ}). This happens because whenever
gluon momentum $q^+$ approaches zero, its energy $q^- =
[m_g^2 + (q^\perp)^2]/q^+$ diverges making $f^{q/g}_{t_r,123}$
vanish exponentially quickly. With $m_g = 0$ it is possible
to approach $q^+ = 0$ without making the gluon energy $q^-$
diverge, for example by keeping $q^\perp = 0$ or by making
$q^\perp$ simultaneously approach 0. Since $f^{q}_{t_r,123}$, $f^{g}_{t_r,123}$
vanishes exponentially quickly it can regulate any fixed,
negative power of $q^+$. The gluon mass is treated as simply
another regularization. We study the $m_g \to 0$ limit is Sec.~\ref{sec:casimir}.

%
\section{\label{sec:rgpep}Effective interactions}
%

The effective Hamiltonian, $\cH_t$ is defined by the following
set of equations:
\beq
\cH_t \es H_0 \ ,
\\
\frac{d}{dt} \cH_t
\es
\left[ \cG_t, \cH_t \right]
 ,
\label{eq:RGPEP}
\eeq
where $H_0$ is the initial Hamiltonian with local interactions,
and $\cG_t$ is the generator of the unitary transformation.
We choose the generator in the following form,
\beq
\cG_t \es \left[ \cH_f, \cH_t \right] \rs -i\pd_f^- \cH_t \ ,
\eeq
where $\cH_f = H_{\psi^2 + G^2}$ is the free Hamiltonian, i.e.,
one obtained by putting the coupling constant, $g = 0$.

Equation~(\ref{eq:RGPEP}) is a nonlinear differential equation
on the space of operators, hence, exact solutions are not easy
to find. For QCD one can take advantage of asymptotic freedom
and compute effective Hamiltonians in perturbative expansion
in powers of the coupling constant. At sufficiently high
energy scales, such expansion should reproduce the exact
solution well enough for the effective Hamiltonians to have
eigenspectra close to the exact eigenspectrum. By defining
the interacting Hamiltonian,
\beq
\cH_{It} \es \cH_t - \cH_f \ ,
\eeq
we get,
\beq
\frac{d}{dt} \cH_t
\es
-(i\pd_f^-)^2 \cH_{It}
+ \left[ -i\pd_f^- \cH_{It}, \cH_{It} \right]
 .
\label{eq:RGPEP2}
\eeq
where we used the fact that $\pd_f^- \cH_f = 0$ to replace
$\pd_f^- \cH_t$ with $\pd_f^- \cH_{It}$. Furthermore, we
introduce the reduced Hamiltonian, $h_t$ through,
\beq
\cH_{It}
\es
e^{-t(i\pd_f^-)^2} h_t
\ .
\label{eq:ht}
\eeq
The RGPEP equation becomes,
\beq
\frac{d}{dt} h_t
\es
e^{t(i\pd_f^-)^2}
\left[ -i\pd_f^- \cH_{It}, \cH_{It} \right] .
\eeq
We expand in powers of the coupling constant $g$,
\beq
h_t \es h_{t , 1} + h_{t , 2} + h_{t , 3} + \dots \ ,
\eeq
where $h_{t,n} \sim g^n$. We arrive at a set
of equations, order by order,
\beq
\frac{d h_{t , 1}}{dt} \es 0 \ ,
\label{eq:RGPEP1st}
\\
\frac{d h_{t , 2}}{dt}
\es
e^{t(i\pd_f^-)^2}
\left[ -i\pd_f^- \cH_{t,1}, \cH_{t,1} \right]
 ,
\label{eq:RGPEP2nd}
\\
\frac{d h_{t , 3}}{dt}
\es
e^{t(i\pd_f^-)^2}
\left[ -i\pd_f^- \cH_{t,1}, \cH_{t,2} \right]
+
e^{t(i\pd_f^-)^2}
\left[ -i\pd_f^- \cH_{t,2}, \cH_{t,1} \right]
 ,
\quad\text{etc.,}
\label{eq:RGPEP3rd}
\eeq
where $\cH_{t,k} = e^{-t(i\pd_f^-)^2} h_{t,k}$. In the
remainder, we solve these equations up to second order.

\subsection{First order}

The first-order RGPEP Eq.~(\ref{eq:RGPEP1st}) has a simple solution,
\beq
h_{t,1}
\es
h_{0,1}
\rs
H_{\psi^2 G} + H_{G^3}
\ .
\label{eq:ht1QCD}
\eeq
Therefore,
\beq
\cH_{t,1}
\es
g
\int[q_1 q_2 q_3]
\,\tdelta_{123}
\, f^q_{t + t_r,123}
\,\cN\left[
\bar\Psi(-q_1) \gamma_\mu T^a \Psi(q_2) G^{\mu a}(q_3)
\right]
\np
\frac{g}{3!}
\int[q_1 q_2 q_3]
\,\tdelta_{123}
\, f^g_{t + t_r,123}
F_{\alpha\beta\gamma}^{abc}(q_1, q_2, q_3)
\, G^{\alpha a}(q_1) G^{\beta b}(q_2) G^{\gamma c}(q_3)
 .
\eeq
The interactions acquire an effective form factor $f_{t,123}$,
which is combined with the regulating factor $f_{t_r,123}
f_{t,123} = f_{t + t_r,123}$. As in the case of Yukawa theory,
the first order solution lowers the energy cutoff from
$1/\sqrt{t_r}$ at $t = 0$ to $1/\sqrt{t + t_r}$ at $t > 0$.
The matrix elements of $\cH_{t,1}$ are insensitive to $t_r$
in the limit $t_r \to 0^+$.

\subsection{Second order}

The second-order calculation is slightly more complicated
than in Yukawa theory~\cite{Serafin:2025ouo} because there are two first-order,
three-leg vertices. The two copies of $\cH_{t,1}$ in the
commutator in Eq.~(\ref{eq:RGPEP2nd}) lead to four terms,
\beq
\frac{d}{dt} h_{t,2}
\es
\int[q_1 q_2 q_3 q_4 q_5 q_6]
\, \tdelta_{123}
\, \tdelta_{456}
\left( \cA'_t + \cB'_t + \cC'_t + \cD'_t \right) ,
\label{eq:dhdt2ndQCD}
\eeq
where
\beq
\cA_t'
\es
A_{t,123.456}^{q.q}
f_{t_r,123}^q
\tilde J_{q\,\mu}^a(q_1, q_2)
G^{\mu a}(q_3)
f_{t_r,456}^q
\tilde J_{q\,\nu}^b(q_4, q_5)
G^{\nu b}(q_6)
\ ,
\\
\cB_t'
\es
A_{t,123.456}^{q.g}
f_{t_r,123}^q
\tilde J_{q\,\mu}^a(q_1, q_2)
G^{\mu a}(q_3)
f_{t_r,456}^g
\tilde J_{g\,\nu}^b(q_4, q_5)
G^{\nu b}(q_6)
\ ,
\\
\cC_t'
\es
A_{t,123.456}^{g.q}
f_{t_r,123}^g
\tilde J_{g\,\mu}^a(q_1, q_2)
G^{\mu a}(q_3)
f_{t_r,456}^q
\tilde J_{q\,\nu}^b(q_4, q_5)
G^{\nu b}(q_6)
\ ,
\\
\cD_t'
\es
A_{t,123.456}^{g.g}
f_{t_r,123}^g
\tilde J_{g\,\mu}^a(q_1, q_2)
G^{\mu a}(q_3)
f_{t_r,456}^g
\tilde J_{g\,\nu}^b(q_4, q_5)
G^{\nu b}(q_6)
\ ,
\eeq
and
\beq
A^{I.J}_{123.456}
\es
\left( - Q^I_{123} + Q^J_{456} \right)
\frac{ f^I_{t, 123} f^J_{t, 456} }{ f^{I.J}_{t,123456} }
\ ,
\eeq
for $I, J \in \{ q, g \}$, where
\beq
Q^q_{ijk}
\es
  \left.q_i^-\right|_{m}
+ \left.q_j^-\right|_{m}
+ \left.q_k^-\right|_{m_g} \ ,
\\
Q^g_{ijk}
\es
  \left.q_i^-\right|_{m_g}
+ \left.q_j^-\right|_{m_g}
+ \left.q_k^-\right|_{m_g} \ ,
\\
f^J_{t, ijk}
\es
e^{-t \left( Q^J_{ijk} \right)^2} \ ,
\\
f^{I.J}_{t, 123456}
\es
e^{-t \left( Q^I_{123} + Q^J_{456} \right)^2} \ .
\eeq

The difference between various functions $A^{I.J}_{123.456}$
for various $I$ and $J$ is entirely contained in the difference
between masses that enter the definitions of the minus
components of various fourvectors $q_i$.

After integrating Eq.~(\ref{eq:dhdt2ndQCD}) over $t$ we get,
\beq
h_{t,2}
\es
h_{0,2}
+
\int[q_1 q_2 q_3 q_4 q_5 q_6]
\, \tdelta_{123}
\, \tdelta_{456}
\left( \cA_t + \cB_t + \cC_t + \cD_t \right) ,
\label{eq:ht2ndQCD}
\eeq
where
\beq
\cA_t
\es
\int_0^t d\tau \cA_\tau'
\ ,
\quad
\text{etc.}
\eeq
Therefore,
\beq
\cA_t
\es
B_{t,123.456}^{q.q}
f_{t_r,123}^q
\tilde J_{q\,\mu}^a(q_1, q_2)
G^{\mu a}(q_3)
f_{t_r,456}^q
\tilde J_{q\,\nu}^b(q_4, q_5)
G^{\nu b}(q_6)
\ ,
\\
\cB_t
\es
B_{t,123.456}^{q.g}
f_{t_r,123}^q
\tilde J_{q\,\mu}^a(q_1, q_2)
G^{\mu a}(q_3)
f_{t_r,456}^g
\tilde J_{g\,\nu}^b(q_4, q_5)
G^{\nu b}(q_6)
\ ,
\\
\cC_t
\es
B_{t,123.456}^{g.q}
f_{t_r,123}^g
\tilde J_{g\,\mu}^a(q_1, q_2)
G^{\mu a}(q_3)
f_{t_r,456}^q
\tilde J_{q\,\nu}^b(q_4, q_5)
G^{\nu b}(q_6)
\ ,
\\
\cD_t
\es
B_{t,123.456}^{g.g}
f_{t_r,123}^g
\tilde J_{g\,\mu}^a(q_1, q_2)
G^{\mu a}(q_3)
f_{t_r,456}^g
\tilde J_{g\,\nu}^b(q_4, q_5)
G^{\nu b}(q_6)
\ ,
\eeq
where
\beq
B^{I.J}_{123.456}
\es
\frac{1}{2} \left( \frac{1}{Q^I_{123}} - \frac{1}{Q^J_{456}} \right)
\left( \frac{ f^I_{t, 123} f^J_{t, 456} }{ f^{I.J}_{t,123456} } - 1 \right)
 ,
\eeq
for $I, J \in \{ q, g \}$.


\begin{table}
\begin{tabular}{c|cccccc}
$I$ & $a$--$g$ & $j$ & $k$ & $l$ & $m$ & $n$ \\
\hline
$S_I$ & 1 & 3 & 3 & 9 & 18 & 6
\end{tabular}
\caption{\label{tab:SI}Symmetry factor $S_I$ for
diagram $D_I$.}
\end{table}

Using Wick's theorem $\cA_t$, $\cB_t$, $\cC_t$, and $\cD_t$
can be evaluated further. Equation~(\ref{eq:ht2ndQCD}) will
be still correct if we make the following substitutions:
\beq
\cA_t
& \to &
g^2
f_{t_r,123}^q f_{t_r,456}^q
\, B_{t,123.456}^{q.q}
\, (D_a + D_b + D_c + D_d + D_e + D_f + D_g) \ ,
\label{eq:cAtoabc}
\\
\cB_t
& \to &
g^2
f_{t_r,123}^q f_{t_r,456}^g
\, B_{t,123.456}^{q.g}
\, S_j D_j \ ,
\label{eq:cBtoabc}
\\
\cC_t
& \to &
g^2
f_{t_r,123}^g f_{t_r,456}^q
\, B_{t,123.456}^{g.q}
\, S_k D_k \ ,
\label{eq:cCtoabc}
\\
\cD_t
& \to &
g^2
f_{t_r,123}^g f_{t_r,456}^g
\, B_{t,123.456}^{g.g}
\, (S_l D_l + S_m D_m + S_n D_n) \ ,
\label{eq:cDtoabc}
\eeq
where $D_a, \dots, D_n$ represent Wick's contractions,
are depicted in Figs.~\ref{fig:wickA}, \ref{fig:wickBC},
\ref{fig:wickDE}, \ref{fig:wickF}, \ref{fig:wickG},
\ref{fig:wickJK}, and \ref{fig:wickLMN}, and are
discussed in Sec.~\ref{sec:quark} and Sec.~\ref{sec:gluon}.
Formulas~(\ref{eq:cAtoabc})--(\ref{eq:cDtoabc}) are correct
only under the integration sign of Eq.~(\ref{eq:ht2ndQCD}).
$\cB_t$, $\cC_t$, and $\cD_t$ contain symmetry factors
$S_I$, see Table~\ref{tab:SI}. In $\cA_t$ all symmetry
factors are equal one and are omitted. For example, $S_j = 3$,
because there are 3 similar diagrams that only differ
by which of the three gluon legs of the right vertex
is contracted with the unique gluon leg of the left
vertex in Fig.~\ref{fig:wickJK}.
Since $F_{\alpha\beta\gamma}^{abc}(q_1, q_2, q_3)$ is
symmetric with respect to simultaneous permutations
$1 \to 2 \to 3 \to 1$, $a \to b \to c \to a$, and
$\alpha \to \beta \to \gamma \to \alpha$, as well as
simultaneous permutations $1 \leftrightarrow 2$,
$a \leftrightarrow b$, and $\alpha\leftrightarrow\beta$,
the three-gluon vertex is symmetric with respect to
permutations of its legs and the three diagrams give
the same result. Hence, instead of computing them
separately, we can compute only one of them and multiply
the result by 3. In another example, $S_m = 18$ because
there are 3 choices of which two legs in diagram $D_m$
should be contracted in the left vertex, 3 choices of
which two legs should be contracted in the right vertex,
and once these are chosen one still has 2 choices
of how to contract two legs with two legs.

The reduced Hamiltonian becomes
\beq
h_{t,2}
\es
h_{0,2}
+
\sum_{J \in \{a, \dots, n\}} h^{(J)}_{t,2} \ .
\eeq
where $h^{(J)}_{t,2}$ for $J \in \{a, \dots, n\}$ come
from Wick's contractions. $h_{0,2}$ contains the bare
interaction vertices, meaning initial vertices at $t = 0$,
and counterterms necessary to renormalize the theory,
\beq
h_{0,2} \es H_{G^4} + H_{\psi^2 G^2} + H_{JJ} + X \ ,
\eeq
where $X$ stands for the counterterms.

We discuss terms involving only quark-gluon vertices,
i.e., diagrams $D_a$ through $D_g$, in Sec.~\ref{sec:quark}.
We discuss terms involving at least one three-gluon vertex,
i.e., diagrams $D_j$ through $D_n$, in Sec.~\ref{sec:gluon}.

\subsection{\label{sec:quark}Wick's contractions involving only quark-gluon vertices}

\begin{figure}[h]
 \centering
 \includegraphics{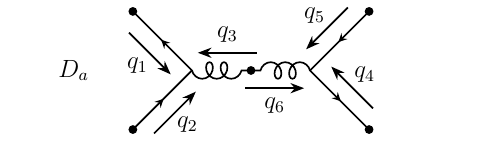}
 \caption{Wick's diagram $D_a$.}
 \label{fig:wickA}
\end{figure}

\begin{figure}[h]
 \centering
 \includegraphics{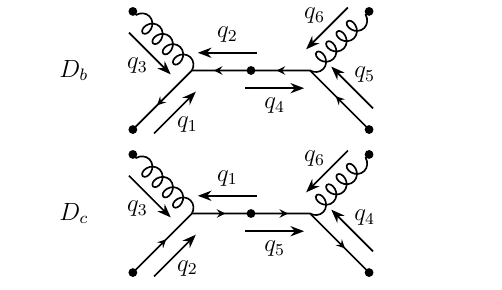}
 \caption{Wick's diagrams $D_b$ and $D_c$.}
 \label{fig:wickBC}
\end{figure}

\begin{figure}[h]
 \centering
 \includegraphics{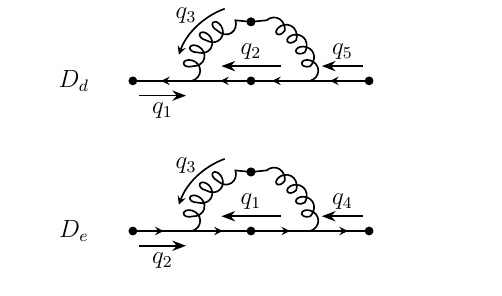}
 \caption{Wick's diagrams $D_d$ and $D_e$.}
 \label{fig:wickDE}
\end{figure}

\begin{figure}[h]
 \centering
 \includegraphics{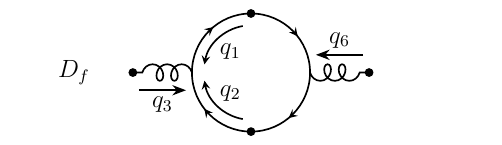}
 \caption{Wick's diagram $D_f$.}
 \label{fig:wickF}
\end{figure}

\begin{figure}[h]
 \centering
 \includegraphics{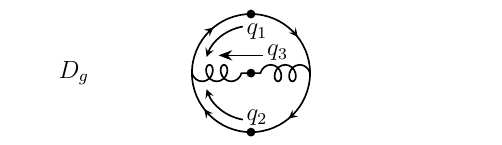}
 \caption{Wick's diagram $D_g$.}
 \label{fig:wickG}
\end{figure}

Diagrams $D_a$ through $D_g$, depicted in
Figs.~\ref{fig:wickA}--\ref{fig:wickG}, involve only
quark-gluon vertices. These terms are analogous to
the diagrams in Figs.~4(a)--(g) in the Yukawa paper.
As compared to Yukawa paper, here the vertices involve
Dirac gamma matrices, $\gamma^\mu$ or $\gamma^\nu$,
color matrices, $T^a$ or $T^b$, and gluons carry
a relativistic index, $\mu$ or $\nu$, and a color
index $a$ or $b$. These differences come from the fact
that quarks possess color charge, and gluons are vector
bosons associated with the gauge group, as opposed to
simple fermions and scalar bosons considered in the Yukawa
paper. For the purpose of computing the effective
Hamiltonians the differences are merely technical
complications. Yukawa paper explains in much more detail
the steps of the calculation. We summarize the results
and recapitulate the most important comments.

Diagrams in Figs.~\ref{fig:wickA} through \ref{fig:wickG}
represent the following Wick's contractions:
\beq
D_a
\es
\cN\left[
\wick{
\bar\Psi(-q_1) \gamma_\mu T^a \Psi(q_2) \c{G}^{\mu a}(q_3)
\bar\Psi(-q_4) \gamma_\nu T^b \Psi(q_5) \c{G}^{\nu b}(q_6)
}
\right] ,
\\
D_b
\es
\cN\left[
\wick{
\bar\Psi(-q_1) \gamma_\mu T^a \c{\Psi}(q_2) G^{\mu a}(q_3)
\c{\bar\Psi}(-q_4) \gamma_\nu T^b \Psi(q_5) G^{\nu b}(q_6)
}
\right] ,
\\
D_c
\es
\cN\left[
\wick{
\c{\bar\Psi}(-q_1) \gamma_\mu T^a \Psi(q_2) G^{\mu a}(q_3)
\bar\Psi(-q_4) \gamma_\nu T^b \c{\Psi}(q_5) G^{\nu b}(q_6)
}
\right] ,
\\
D_d
\es
\cN\left[
\wick{
\bar\Psi(-q_1) \gamma_\mu T^a \c1{\Psi}(q_2) \c2{G}^{\mu a}(q_3)
\c1{\bar\Psi}(-q_4) \gamma_\nu T^b \Psi(q_5) \c2{G}^{\nu b}(q_6)
}
\right] ,
\\
D_e
\es
\cN\left[
\wick{
\c1{\bar\Psi}(-q_1) \gamma_\mu T^a \Psi(q_2) \c2{G}^{\mu a}(q_3)
\bar\Psi(-q_4) \gamma_\nu T^b \c1{\Psi}(q_5) \c2{G}^{\nu b}(q_6)
}
\right] ,
\\
D_f
\es
\cN\left[
\wick{
\c1{\bar\Psi}(-q_1) \gamma_\mu T^a \c2{\Psi}(q_2) G^{\mu a}(q_3)
\c2{\bar\Psi}(-q_4) \gamma_\nu T^b \c1{\Psi}(q_5) G^{\nu b}(q_6)
}
\right] ,
\\
D_g
\es
\cN\left[
\wick{
\c1{\bar\Psi}(-q_1) \gamma_\mu T^a \c2{\Psi}(q_2) \c3{G}^{\mu a}(q_3)
\c2{\bar\Psi}(-q_4) \gamma_\nu T^b \c1{\Psi}(q_5) \c3{G}^{\nu b}(q_6)
}
\right] .
\eeq
The contractions are
\beq
\wick{
\c{\bar\Psi}_{a_1 c_1}(-q_1)
\,
\c{\Psi}_{a_2 c_2}(q_2)
}
\es
\frac{ \theta(q_1^+) }{ |q_1^+| }
\, \tdelta(q_1 + q_2)
\, \delta_{c_2, c_1}
\, (\slashed q_1 - m)_{a_2, a_1}
\ ,
\label{eq:contractionPsiPsi1QCD}
\\
\wick{
\c{\Psi}_{a_1 c_1}(q_1)
\,
\c{\bar\Psi}_{a_2 c_2}(-q_2)
}
\es
\frac{ \theta(q_1^+) }{|q_1^+|}
\, \tdelta(q_1 + q_2)
\, \delta_{c_1, c_2}
\, (\slashed q_1 + m)_{a_1, a_2}
\ ,
\label{eq:contractionPsiPsi2QCD}
\\
\wick{
\c{G}^{\mu \, a}(q_1)
\,
\c{G}^{\nu \, b}(q_2)
}
\es
\frac{ \theta(q_1^+) }{ |q_1^+| }
\,\tdelta(q_1 + q_2)
\,\delta_{ab}
\,d^{\mu\nu}(q_1)
\ ,
\label{eq:contractionGG}
\eeq
where indices $a_1, a_2 = 1, 2, 3, 4$ enumerate entries
in the spinor space, while indices $c_1, c_2 = 1, 2, 3$
enumerate entries in the color space of the quark field.
The polarization tensor is
\beq
d^{\mu\nu}(q)
\es
- g^{\mu\nu}
+ \frac{n^\mu q^\nu + n^\nu q^\mu}{q^+}
- n^\mu n^\nu \frac{q^2}{(q^+)^2}
\ ,
\label{eq:dmunu}
\eeq
Note that $q^- = [m_g^2 + (q^\perp)^2]/q^+$. Therefore,
$q^2 = m_g^2$, and $d^{\mu\nu}(q)$ depends only on $q^+$
and $q^\perp$, and does not depend on $m_g$. Explicitly,
\beq
d^{--}(q)
\es
\frac{4 (q^\perp)^2}{(q^+)^2}
\ ,
\\
d^{ij}(q)
\es
  \delta^{ij}
\ ,
\\
d^{-i}(q)
\es
d^{i-}(q)
\rs
\frac{2 q^i}{q^+}
\ ,
\\
d^{+\nu}(q)
\es
d^{\nu+}(q)
\rs
0
\ .
\eeq

For $J \in \{a, \dots, g\}$
\beq
h_{t,2}^{(J)}
\es
g^2
\int[q_1 q_2 q_3 q_4 q_5 q_6]
\, \tdelta_{123}
\, \tdelta_{456}
f_{t_r,123}^q
f_{t_r,456}^q
B_{t,123.456}^{q.q}
D_J
\ .
\eeq

Diagram $D_a$, Fig.~\ref{fig:wickA} gives,
\beq
h_{t,2}^{(a)}
\es
g^2
\int[q_1 q_2 q_3 q_4 q_5]
\, \tdelta_{123}
\, \tdelta_{45.3}
f_{t_r,123}^q
f_{t_r,45.3}^q
B_{t,123.45(-3)}^{q.q}
\frac{ \theta(q_3^+) }{ q_3^+ }
\nt
\,d^{\mu\nu}(q_3)
\cN\left[
\bar\Psi(-q_1) \gamma_\mu T^a \Psi(q_2)
\bar\Psi(-q_4) \gamma_\nu T^a \Psi(q_5)
\right]
 .
\eeq

Diagrams $D_b$ and $D_c$, Fig.~\ref{fig:wickBC} give,
\beq
h_{t,2}^{(b)} + h_{t,2}^{(c)}
\es
g^2
\int[q_1 q_2 q_3 q_5 q_6]
\, \tdelta_{123}
\, \tdelta_{56.2}
f_{t_r,123}^q
f_{t_r,(-2)56}^q
B_{t,123.(-2)56}^{q.q}
\frac{ 1 }{ q_2^+ }
\nt
\cN\left[
\bar\Psi(-q_1) \gamma_\mu T^a G^{\mu a}(q_3)
\, (\slashed q_2 + m)
\gamma_\nu T^b G^{\nu b}(q_6) \Psi(q_5)
\right]
 .
\eeq

Diagrams $D_d$ and $D_e$, Fig.~\ref{fig:wickDE} give,
\beq
h_{t,2}^{(d)} + h_{t,2}^{(e)}
\es
C_F g^2
\int[q_1 q_2 q_3]
\, \tdelta_{23.1}
\frac{ B_{t,(-1)23.1(-2)(-3)}^{q.q} }{ |q_2^+ q_3^+| }
(f_{t_r,(-1)23}^q)^2
\nt
d^{\mu\nu}(q_3)
\cN\left[
\bar\Psi(q_1) \gamma_\mu
\, (\slashed q_2 + m)
\gamma_\nu \Psi(q_1)
\right]
\left[
  \theta(q_2^+) \theta(q_3^+)
- \theta(-q_2^+) \theta(-q_3^+)
\right]
 ,
\nn
\label{eq:ht2de}
\eeq
where
\beq
B_{t,(-1)23.1(-2)(-3)}^{q.q}
\es
\frac{ (f_{t,(-1)23}^q)^2 - 1 }
{ \left. q_2^- \right|_{m} + \left. q_3^- \right|_{m_g} - \left. q_1^-\right|_{m} }
\ ,
\label{eq:BqqDde}
\eeq
and $C_F = (N_c^2 - 1)/(2 N_c) = 4/3$ for the number of colors
$N_c = 3$. Using the Gordon identity,
\beq
h_{t,2}^{(d)} + h_{t,2}^{(e)}
\es
  \int[q]\,
  \frac{\delta\tilde m^2_{t,2}}{q^+}
  \cN\left[
  \bar\Psi(q)
  \frac{\gamma^+}{2}
  \Psi(q)
  \right]
 ,
\eeq
where
\beq
\delta\tilde m^2_{t,2}
\es
C_F g^2
\int[q_2 q_3]
\, \tdelta_{23.q}
\frac{ B_{t,(-q)23.q(-2)(-3)}^{q.q} }{ q_2^+ q_3^+ }
(f_{t_r,(-q)23}^q)^2
\theta(q_2^+) \theta(q_3^+)
\nt
d_{\mu\nu}(q_3)
  \bar u_{\sigma}(q)
\gamma^\mu
\, (\slashed q_2 + m)
\gamma^\nu
  u_{\sigma}(q)
\ .
\label{eq:dtm2t2}
\eeq

Diagram $D_f$, Fig.~\ref{fig:wickF} gives,
\beq
h_{t,2}^{(f)}
\es
\int[q_3]
\theta(q_3^+)
\tilde\Pi^{q}_{\mu\nu}(q_3)
\cN\left[
G^{\mu a}(-q_3)
G^{\nu a}(q_3)
\right] ,
\label{eq:ht2f}
\eeq
where
\beq
\tilde\Pi^{q}_{\mu\nu}(q_3)
\es
g^2
\int[q_1 q_2]
\, \tdelta_{12.3}
\frac{ \theta(q_1^+) }{ |q_1^+| }
\frac{ \theta(q_2^+) }{ |q_2^+| }
B_{t,12(-3).(-1)(-2)3}^{q.q}
(f_{t_r,12(-3)}^q)^2
\nt
T_f
\mathop{\mathrm{Tr}}\left[
\gamma_\mu
\, (\slashed q_2 + m)
\gamma_\nu
\, (\slashed q_1 - m) \right]
 ,
\label{eq:PimunuQ}
\eeq
with $T_f = \frac{1}{2}$, and
\beq
B_{t,12(-3).(-1)(-2)3}^{q.q}
\es
\frac{ (f^q_{t, 12(-3)})^2 - 1 }{ q_1^- + q_2^- - q_3^- }
\ .
\label{eq:BqqDf}
\eeq
$\tilde\Pi^{q}_{\mu\nu}(q_3)$ is to some extent similar
to vacuum polarization tensor as defined in calculations
of correlation functions, but should not be confused
with it. $\tilde\Pi^{q}_{\mu\nu}(q_3)$ is just a part of
the expression for the effective Hamiltonian, and as such
it will be an ingredient of a calculation of the effective
vacuum polarization. Only components $\mu, \nu = -, 1, 2$
contribute to Eq.~(\ref{eq:ht2f}), because $G^{+ \, a}(q) = 0$
due to gauge choice, and they satisfy relations,
\beq
\tilde\Pi^{q}_{-i}(q_3)
\es
\tilde\Pi^{q}_{i-}(q_3)
\rs
- \tilde\Pi^{q}_{--}(q_3) \frac{2 q_3^i}{q_3^+}
\ ,
\label{eq:PiMinusIQ}
\\
\tilde\Pi^{q}_{ij}(q_3)
\es
  \tilde\Pi^{q}_{--}(q_3) \frac{4 q_3^i q_3^j}{(q_3^+)^2}
+ \delta_{ij} \delta\tilde\mu_{t,q,2}^2(q_3)
\ ,
\label{eq:PiIJQ}
\eeq
where
\beq
\tilde\Pi^{q}_{--}(q_3)
\es
T_f
g^2
\int[q_1 q_2]
\frac{ \theta(q_1^+) }{ |q_1^+| }
\frac{ \theta(q_2^+) }{ |q_2^+| }
\, \tdelta_{12.3}
B_{t,12(-3).(-1)(-2)3}^{q.q}
(f_{t_r,12(-3)}^q)^2
\cdot
2 q_1^+ q_2^+
\ ,
\\
\delta\tilde\mu_{t,q,2}^2(q_3)
\es
2 T_f g^2
\int[q_1 q_2]
\frac{ \theta(q_1^+) }{ |q_1^+| }
\frac{ \theta(q_2^+) }{ |q_2^+| }
\, \tdelta_{12.3}
B_{t,12(-3).(-1)(-2)3}^{q.q}
(f_{t_r,12(-3)}^q)^2
\left[ \cM_{12}^2 - 2 k^2 \right] ,
\nn
\label{eq:delTildMuQ}
\eeq
with $\cM_{12}^2 = \frac{m^2 + k^2}{x_1 x_2}$.
$k^\perp = (k^1, k^2)$ and $x_1 = q_1^+/q_3^+ = 1 - x_2$,
are relative momentum variables, such that
\beq
q_1^i \es k^i + x_1 q_3^i \ ,
\label{eq:fermRel1}
\\
q_2^i \es -k^i + x_2 q_3^i \ .
\label{eq:fermRel2}
\eeq
For simplicity, $k^2 = (k^\perp)^2$, not to be confused
with $k^2$ as the second coordinate of $k^\perp$.
Relations (\ref{eq:PiMinusIQ}) and (\ref{eq:PiIJQ})
can be checked explicitly using,
\beq
\mathop{\mathrm{Tr}}\left[
\gamma_\mu
\, (\slashed q_2 + m)
\gamma_\nu
\, (\slashed q_1 - m) \right]
\es
  4 q_{1\mu} q_{2\nu}
+ 4 q_{1\nu} q_{2\mu}
- 4 g_{\mu\nu} ( q_1 \cdot q_2 + m^2 )
\ .
\label{eq:spinorTrace}
\eeq
Any term that is linear in $k^1$, $k^2$, or $k^1 k^2$,
integrates to zero in Eq.~(\ref{eq:PimunuQ}), because
all other factors of the integrand depend on $(k^\perp)^2$.
In particular this means that $k^i k^j$ can be replaced
under the integral with $\delta_{ij} (k^\perp)^2/2$,
which leads to Eq.~(\ref{eq:PiIJQ}).

Since $q_\mu G^{\mu \, a}(q) = 0$,
\beq
G^{- \, a}(q)
\es
\frac{2 q^j G^{j \, a}(q)}{q^+}
\ ,
\eeq
and one can express Eq.~(\ref{eq:PimunuQ}) in terms of
$G^{i \, a}(q)$ only. Using Eqs.~(\ref{eq:PiMinusIQ})
and (\ref{eq:PiIJQ}), we obtain
\beq
h_{t,2}^{(f)}
\es
\int[q_3]
\theta(q_3^+)
\delta\tilde\mu_{t,q,2}^2(q_3)
\cN\left[
G^{i a}(-q_3)
G^{i a}(q_3)
\right] ,
\label{eq:ht2fsecond}
\eeq
where only $\delta\tilde\mu_{t,q,2}^2(q_3)$ contributes,
while all instances of $\tilde\Pi_{--}^q(q_3)$
canceled each other. Comparison with Eq.~(\ref{eq:freeCan})
reveals that $\delta\tilde\mu_{t,q,2}^2(q_3)$ is a correction
to the free gluon mass $m_g$.

Finally, diagram $D_g$, Fig.~\ref{fig:wickG} gives,
\beq
h_{t,2}^{(g)} \es 0 \ ,
\eeq
due to the small longitudinal momentum cutoff.

\subsection{\label{sec:gluon}Wick's contractions involving triple gluon vertices}

\begin{figure}[h]
 \centering
 \includegraphics{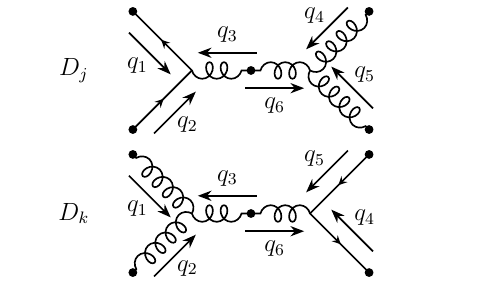}
 \caption{Wick's diagrams $D_j$ and $D_k$.}
 \label{fig:wickJK}
\end{figure}

Diagram $D_j$, Fig.~\ref{fig:wickJK} gives,
\beq
h_{t,2}^{(j)}
\es
g^2
\int[q_1 q_2 q_3 q_4 q_5 q_6]
\, \tdelta_{123}
\, \tdelta_{456}
f_{t_r,123}^q
f_{t_r,456}^g
B_{t,123.456}^{q.g}
\cdot 3 D_j
\ ,
\eeq
where
\beq
D_j
\es
\frac{1}{3!}
F_{\alpha\beta\gamma}^{abc}(q_4, q_5, q_6)
\,\cN\left[
\wick{
\bar\Psi(-q_1) \gamma_\mu T^d \Psi(q_2) \c{G}^{\mu d}(q_3)
\, G^{\alpha a}(q_4) G^{\beta b}(q_5) \c{G}^{\gamma c}(q_6)
}
\right]
 .
\eeq
Therefore,
\beq
h_{t,2}^{(j)}
\es
g^2
\int[q_1 q_2 q_3 q_4 q_5]
\, \tdelta_{123}
\, \tdelta_{45.3}
f_{t_r,123}^q
f_{t_r,45(-3)}^g
B_{t,123.45(-3)}^{q.g}
\frac{ \theta(q_3^+) }{ |q_3^+| }
\nt
d^{\mu\gamma}(q_3)
\frac{1}{2}
F_{\alpha\beta\gamma}^{abc}(q_4, q_5, -q_3)
\,\cN\left[
\bar\Psi(-q_1) \gamma_\mu T^c \Psi(q_2)
\, G^{\alpha a}(q_4) G^{\beta b}(q_5)
\right]
 .
\eeq
Similarly, diagram $D_k$, Fig.~\ref{fig:wickJK} gives,
\beq
h_{t,2}^{(k)}
\es
g^2
\int[q_1 q_2 q_3 q_4 q_5 q_6]
\, \tdelta_{123}
\, \tdelta_{456}
f_{t_r,123}^g
f_{t_r,456}^q
B_{t,123.456}^{g.q}
\cdot 3 D_k
\ ,
\eeq
where
\beq
D_k
\es
\frac{1}{3!}
F_{\alpha\beta\gamma}^{abc}(q_1, q_2, q_3)
\,\cN\left[
\wick{
G^{\alpha a}(q_1) G^{\beta b}(q_2) \c{G}^{\gamma c}(q_3)
\,\bar\Psi(-q_4) \gamma_\mu T^d \Psi(q_5) \c{G}^{\mu d}(q_6)
}
\right] .
\eeq
Therefore,
\beq
h_{t,2}^{(k)}
\es
g^2
\int[q_1 q_2 q_3 q_4 q_5]
\, \tdelta_{123}
\, \tdelta_{45.3}
f_{t_r,123}^g
f_{t_r,45(-3)}^q
B_{t,123.45(-3)}^{g.q}
\frac{ \theta(q_3^+) }{ |q_3^+| }
\nt
d^{\gamma\mu}(q_3)
\frac{1}{2}
F_{\alpha\beta\gamma}^{abc}(q_1, q_2, q_3)
\,\cN\left[
G^{\alpha a}(q_1) G^{\beta b}(q_2)
\,\bar\Psi(-q_4) \gamma_\mu T^c \Psi(q_5)
\right] .
\eeq
Substituting $q_3 \to -q_3$, relabeling
$1 \leftrightarrow 4$, and $2 \leftrightarrow 5$,
then shifting $G^{\alpha a}(q_4) G^{\beta b}(q_5)$
to the right under the normal ordering sign, finally
using symmetries $d^{\gamma\mu}(-q_3) = d^{\mu\gamma}(q_3)$,
and $B_{t,45(-3).123}^{g.q} = -B_{t,123.45(-3)}^{q.g}$,
$h_{t,2}^{(k)}$ is rewritten in a form almost identical
to that of $h_{t,2}^{(j)}$. The only difference is that
where $h_{t,2}^{(j)}$ has $\theta(q_3^+)$,
$h_{t,2}^{(k)}$ has $-\theta(-q_3^+)$.
Since $\theta(q_3^+) - \theta(-q_3^+)
= \mathop{\mathrm{sign}}(q_3^+)$,
the sum becomes
\beq
h_{t,2}^{(j)} + h_{t,2}^{(k)}
\es
g^2
\int[q_1 q_2 q_3 q_4 q_5]
\, \tdelta_{123}
\, \tdelta_{45.3}
f_{t_r,123}^q
f_{t_r,45(-3)}^g
B_{t,123.45(-3)}^{q.g}
\frac{ 1 }{ q_3^+ }
\nt
d^{\mu\gamma}(q_3)
\frac{1}{2}
F_{\alpha\beta\gamma}^{abc}(q_4, q_5, -q_3)
\,\cN\left[
\bar\Psi(-q_1) \gamma_\mu T^c \Psi(q_2)
\, G^{\alpha a}(q_4) G^{\beta b}(q_5)
\right]
 .
\eeq

For $J \in \{l, m, n\}$
\beq
h_{t,2}^{(J)}
\es
g^2
\int[q_1 q_2 q_3 q_4 q_5 q_6]
\, \tdelta_{123}
\, \tdelta_{456}
f_{t_r,123}^g
f_{t_r,456}^g
B_{t,123.456}^{g.g}
S_J D_J
\ .
\eeq

\begin{figure}[h]
 \centering
 \includegraphics{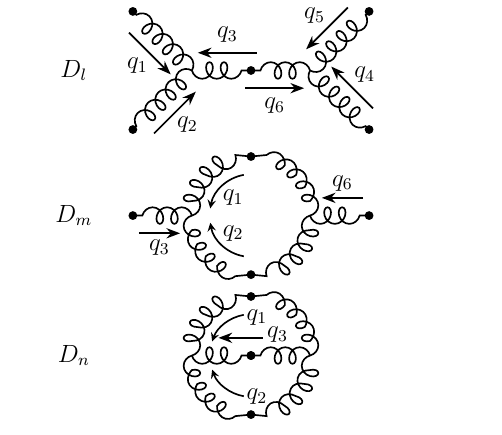}
 \caption{Wick's diagrams $D_l$, $D_m$, and $D_n$.}
 \label{fig:wickLMN}
\end{figure}

Diagrams $D_l$, $D_m$, and $D_n$, Fig.~\ref{fig:wickLMN} are,
\beq
D_l
\es
\frac{1}{3!}
F_{\alpha\beta\gamma}^{abc}(q_1, q_2, q_3)
\frac{1}{3!}
F_{\mu\nu\rho}^{def}(q_4, q_5, q_6)
\nt
\cN\left[
\wick{
G^{\alpha a}(q_1) G^{\beta b}(q_2) \c{G}^{\gamma c}(q_3)
\, G^{\mu d}(q_4) G^{\nu e}(q_5) \c{G}^{\rho f}(q_6)
}
\right] ,
\\
D_m
\es
\frac{1}{3!}
F_{\alpha\beta\gamma}^{abc}(q_1, q_2, q_3)
\frac{1}{3!}
F_{\mu\nu\rho}^{def}(q_4, q_5, q_6)
\nt
\cN\left[
\wick{
\c1{G}^{\alpha a}(q_1) \c2{G}^{\beta b}(q_2) G^{\gamma c}(q_3)
\, \c1{G}^{\mu d}(q_4) \c2{G}^{\nu e}(q_5) G^{\rho f}(q_6)
}
\right] .
\\
D_n
\es
\frac{1}{3!}
F_{\alpha\beta\gamma}^{abc}(q_1, q_2, q_3)
\frac{1}{3!}
F_{\mu\nu\rho}^{def}(q_4, q_5, q_6)
\nt
\cN\left[
\wick{
\c1{G}^{\alpha a}(q_1) \c2{G}^{\beta b}(q_2) \c3{G}^{\gamma c}(q_3)
\, \c1{G}^{\mu d}(q_4) \c2{G}^{\nu e}(q_5) \c3{G}^{\rho f}(q_6)
}
\right] .
\eeq
Therefore,
\beq
h_{t,2}^{(l)}
\es
g^2
\int[q_1 q_2 q_3 q_4 q_5]
\, \tdelta_{123}
\, \tdelta_{45.3}
f_{t_r,123}^g
f_{t_r,45(-3)}^g
B_{t,123.45(-3)}^{g.g}
\frac{ \theta(q_3^+) }{ |q_3^+| }
\nt
\frac{1}{2}
F_{\alpha\beta\gamma}^{abc}(q_1, q_2, q_3)
\frac{1}{2}
F_{\mu\nu\rho}^{dec}(q_4, q_5, -q_3)
\nt
\,d^{\gamma\rho}(q_3)
\cN\left[
G^{\alpha a}(q_1) G^{\beta b}(q_2)
\, G^{\mu d}(q_4) G^{\nu e}(q_5)
\right]
 ,
\\
h_{t,2}^{(m)}
\es
\frac{1}{2}
g^2
\int[q_1 q_2 q_3]
\, \tdelta_{12.3}
f_{t_r,12(-3)}^g
f_{t_r,(-1)(-2)3}^g
B_{t,12(-3).(-1)(-2)3}^{g.g}
\frac{ \theta(q_1^+) }{ |q_1^+| }
\frac{ \theta(q_2^+) }{ |q_2^+| }
\nt
F_{\alpha\beta\gamma}^{abc}(q_1, q_2, -q_3)
F_{\mu\nu\rho}^{abf}(-q_1, -q_2, q_3)
\,d^{\alpha\mu}(q_1)
\,d^{\beta\nu}(q_2)
\cN\left[ G^{\gamma c}(-q_3) G^{\rho f}(q_3) \right]
 ,
\\
h_{t,2}^{(n)} \es 0 \ ,
\eeq
where $h_{t,2}^{(n)}$ is zero because of small longitudinal
momentum cutoff.

Since $f^{abc} f^{abf} = C_A \delta^{cf}$,
where $C_A = N_c$, the gluon mass term can
be written as
\beq
h_{t,2}^{(m)}
\es
\int[q_3]
\theta(q_3^+)
\tilde\Pi^{g}_{\mu\nu}(q_3)
\cN\left[
G^{\mu a}(-q_3)
G^{\nu a}(q_3)
\right] ,
\eeq
where
\beq
\tilde\Pi^{g}_{\mu\nu}(q_3)
\es
\frac{1}{2}
C_A
g^2
\int[q_1 q_2]
\, \tdelta_{12.3}
(f_{t_r,12(-3)}^g)^2
B_{t,12(-3).(-1)(-2)3}^{g.g}
\frac{ \theta(q_1^+) }{ |q_1^+| }
\frac{ \theta(q_2^+) }{ |q_2^+| }
\nt
d^{\alpha\gamma}(q_1)
\left[
  (q_{2\mu} - q_{1\mu}) g_{\alpha\beta}
- (q_{3\alpha} + q_{2\alpha}) g_{\beta\mu}
+ (q_{1\beta} + q_{3\beta}) g_{\mu\alpha}
\right]
\nt
d^{\beta\delta}(q_2)
\left[
  (q_{2\nu} - q_{1\nu}) g_{\gamma\delta}
- (q_{3\gamma} + q_{2\gamma}) g_{\delta\nu}
+ (q_{1\delta} + q_{3\delta}) g_{\nu\gamma}
\right] .
\label{eq:PimunuG}
\eeq
Analogously to $\tilde\Pi_{\mu\nu}^q$, with a use
of Eqs.~(\ref{eq:fermRel1}) and (\ref{eq:fermRel2}),
we find,
\beq
\tilde\Pi^{g}_{-i}(q_3)
\es
\tilde\Pi^{g}_{i-}(q_3)
\rs
- \tilde\Pi^{g}_{--}(q_3) \frac{ 2 q_3^i }{ q_3^+ }
\ .
\label{eq:PiMinusIG}
\\
\tilde\Pi^{g}_{ij}(q_3)
\es
  \tilde\Pi^{g}_{--}(q_3) \frac{ 4 q_3^i q_3^j }{ (q_3^+)^2 }
+ \delta_{ij} \delta\tilde\mu_{t,g,2}^2(q_3)
\ ,
\label{eq:PiIJG}
\eeq
where
\beq
\tilde\Pi^{g}_{--}(q_3)
\es
C_A
g^2
\int[q_1 q_2]
\, \tdelta_{12.3}
(f_{t_r,12(-3)}^g)^2
B_{t,12(-3).(-1)(-2)3}^{g.g}
\frac{ \theta(q_1^+) }{ |q_1^+| }
\frac{ \theta(q_2^+) }{ |q_2^+| }
\frac{(q_2^+ - q_1^+)^2}{4}
\ .
\nn
\\
\delta\tilde\mu_{t,g,2}^2(q_3)
\es
C_A
g^2
\int[q_1 q_2]
\, \tdelta_{12.3}
(f_{t_r,12(-3)}^g)^2
B_{t,12(-3).(-1)(-2)3}^{g.g}
\frac{ \theta(q_1^+) }{ |q_1^+| }
\frac{ \theta(q_2^+) }{ |q_2^+| }
\cdot
2 (k^\perp)^2 \frac{ (1 - x_1 x_2)^2 }{ (x_1 x_2)^2 }
\ .
\nn
\label{eq:delTildMuG}
\eeq
Therefore,
\beq
h_{t,2}^{(m)}
\es
\int[q_3]
\theta(q_3^+)
\delta\tilde\mu_{t,g,2}^2(q_3)
\cN\left[
G^{i a}(-q_3)
G^{i a}(q_3)
\right] ,
\label{eq:ht2m}
\eeq
and $\delta\tilde\mu_{t,g,2}^2(q_3)$ is the second-order
contribution to the gluon mass term in the Hamiltonian
that comes from the gluon loop.

%
\section{\label{sec:counterterms}Counterterms}
%

In this section we study the $t_r \to 0$ limit and find counterterms that make this limit well defined. The rule for finding the counterterms is to ensure that all matrix elements of the effective Hamiltonians are well defined in the $t_r \to 0$ limit. We retain a finite gluon mass while taking $t_r\to 0$ which ensures that all matrix elements, even those outside the color singlet sector, are finite. We then take $m_g\to 0$ in Sec.~\ref{sec:casimir}. In the limit $t_r\to 0$, $m_g\to 0$ we only expect the matrix elements in the color singlet sector to remain finite.

The Hamiltonian of QCD is an unbounded operator that acts on an infinite-dimensional Hilbert space. Such an operator cannot be well defined on the whole of Hilbert space but at most on a dense subspace of it. Therefore, a mathematically rigorous definition of the Hamiltonian requires us to provide its domain. Simply put, a matrix element is formed from an operator and two states, and so finiteness of matrix elements depends on the set of states one is allowed to consider, i.e. the domain of the operator. Loosely speaking, the Hilbert space in which we work is the Fock space build by quark and gluon creation operators, $b_{p\sigma c}^\dag$, $d_{p\sigma c}^\dag$, and $a_{p\sigma c}^\dag$, on top of the free vacuum state $\ket{0}$. Due to commutation and anticommutation relations between these operators, the states have the proper symmetries required by indistinguishability. 

We define a subspace, $F_0$ of the Fock space to be the space of vectors that have nonzero components in finitely many Fock sectors whose wave functions have compact support assuming that the longitudinal momenta, $p^+$ belong to the open set $(0, \infty)$. This means in particular that
the wave functions are zero for momenta of particles that are larger than some finite number or their
longitudinal momenta are smaller than some positive number. Effectively, longitudinal momentum of each
particle is limited from below by some $\epsilon^+ > 0$, except that $\epsilon^+$ depends now on the wave function in question. The universal, positive $\epsilon^+$ that limited the longitudinal momenta in interaction terms of the Hamiltonian is at this point removed. Henceforth, we choose $F_0$ for the domain of the Hamiltonians.

The ultimate goal is to define a Hamiltonian that is a self-adjoint operator. Often the first step is to define an operator that is
symmetric and then to seek its self-adjoint extensions.
The Hamiltonians that we calculate will be proved to be
well-defined symmetric forms with $F_0$ as the domain,
i.e., their matrix elements exist.
It is reasonable to expect that these forms correspond to
symmetric operators. Finding self-adjoint extensions of
these symmetric operators is a technically challenging
task and we do not attempt it here. Nevertheless, the
statement of the problem itself is valuable and, to our
knowledge, novel. In particular, the problem of zero
modes is reinterpreted as closely related to the problem
of finding self-adjoint extensions of symmetric operators.

A generic state with $R$ quarks, $S$ antiquarks, and
$n - R - S$ gluons can be written as
\beq
\ket{\psi}
\es
\sum_{\text{discrete}}
\int[p_1 \dots p_n]
\psi_\text{discrete}(p_1, \dots, p_n)
\left(
\prod_{k = 1}^n \frac{\theta(p_k^+)}{p_k^+} q_k^\dag
\right)
\ket{0} ,
\nn
\eeq
where, $q_k = b_k$ for $1 \le k \le R$, $q_k = d_k$
for $R+1 \le k \le R+S$, and $q_k = a_k$
for $R+S+1 \le k \le n$.

The contractions of these states with various operators are as follows: 
\beq
\wick{
\c{\Psi}(q) \c{b}_{p \lambda c}^\dag
}
\es
  \tdelta_{q.p}
  \theta_\epsilon(p^+)
  \chi_c
  u_{\lambda}(p)
\ ,
\\
\wick{
\c{\bar\Psi}(-q) \c{d}_{p \lambda c}^\dag
}
\es
  \tdelta_{q.p}
  \theta_\epsilon(p^+)
  \bar v_{\lambda}(p)
  \chi_c^\dag
\ ,
\\
\wick{
\c{G}^{\mu \, a}(q) \c{a}_{p \lambda c}^\dag
}
\es
  \tdelta_{q.p}
  \theta_\epsilon(p^+)
  \delta_{ac}
  \varepsilon_\lambda^\mu(p)
\ ,
\\
\wick{
\c{b}_{p \lambda c}
\c{\bar\Psi}(-q)
}
\es
\tdelta_{qp}
\theta_\epsilon(p^+)
\bar u_{\lambda}(p)
\chi_c^\dag
\ ,
\\
\wick{
\c{d}_{p \lambda c}
\c{\Psi}(q)
}
\es
\delta_{qp}
\theta_\epsilon(p^+)
\chi_c
v_{\lambda}(p)
\ ,
\\
\wick{
\c{a}_{p \lambda c} \c{G}^{\mu \, a}(q)
}
\es
\tdelta_{qp}
\theta_\epsilon(p^+)
\delta_{ac}
\varepsilon_\lambda^{\mu*}(p)
\ ,
\eeq
where
\beq
\varepsilon_\sigma^i(q)
\es
\varepsilon_\sigma^i
\ ,
\\
\varepsilon_\sigma^-(q)
\es
\frac{ 2 q^i \varepsilon_\sigma^i }{ q^+ }
\ ,
\eeq
\beq
u_\sigma(p)
\es
\begin{bmatrix}
\sqrt{p^+} \xi_\sigma
\\
\frac{i \sigma^\perp \times p^\perp + m}{\sqrt{p^+}} \xi_\sigma
\end{bmatrix}
,
\eeq
and $\sigma^\perp \times p^\perp = \sigma^1 p^2
- \sigma^2 p^1$.

One can see that contractions between Hamiltonian terms and states introduce spinors or polarization vectors and potentially some singular $1/q^+$ or $1/\sqrt{q^+}$ factors, but these singularities, so called endpoint singularities, are regulated by the compactness of the wave functions. In particular, compactness implies that the support of the integrand is bounded. Therefore, the only way a matrix element can diverge is when the interaction kernel becomes unbounded as $t_r \to 0$. For example, the whole kernel can be shifted to infinity as $t_r$ approaches zero. Another exemplary possibility is that the kernel approaches a function of the form $1/(q^+)^2$, which is not integrable, as $t_r$ approaches zero. Both examples are found in the effective QCD Hamiltonian and we discuss them below. With the
understanding that the wave functions regulate endpoint
singularities we can study the Hamiltonian terms
themselves instead of their matrix elements. Therefore,
our analysis can be slightly simplified.

\begin{figure}[h]
 \centering
 \includegraphics{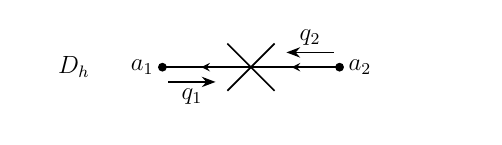}
 \caption{Wick's diagram for the fermion mass counterterm.}
 \label{fig:wickH}
\end{figure}

First, we discuss the quark self-energy terms that arise
from diagrams $D_d$ and $D_e$, Fig.~\ref{fig:wickDE},
Eq.~(\ref{eq:ht2de}). $\delta\tilde m^2_{t,2}$ diverges
as $t_r \to 0$, leading to diverging matrix elements.
Hence, one needs to add a mass counterterm.
$\delta\tilde m^2_{t,2}$ is defined with
$B_{t,(-1)23.1(-2)(-3)}^{q.q}$, Eq.~(\ref{eq:BqqDde})
in the integrand. If the numerator of
$B_{t,(-1)23.1(-2)(-3)}^{q.q}$ were just the form factor,
$(f_{t,(-1)23}^q)^2$, then the integral would be
finite in the $t_r \to 0$ limit. It is the $-1$ term
in the numerator of $B_{t,(-1)23.1(-2)(-3)}^{q.q}$
that leads to a divergent integral. To counter this
divergence we add a counterterm,
\beq
\cH_{t,2}^{(h)}
\es
  \int[q]\,
  \frac{\delta m^2_{t,X,2}}{q^+}
  \cN\left[
  \bar\Psi(q)
  \frac{\gamma^+}{2}
  \Psi(q)
  \right] ,
\eeq
where
\beq
\delta m^2_{t,X,2}
\es
C_F g^2
\int[q_2 q_3]
\, \tdelta_{23.q}
\frac{ \theta(q_2^+) \theta(q_3^+) }{ q_2^+ q_3^+ }
\frac{ (f_{t_r,(-q)23}^q)^2 }
{ \left. q_2^- \right|_{m} + \left. q_3^- \right|_{m_g}
- \left. q_1^-\right|_{m} }
\nt
d_{\mu\nu}(q_3)
  \bar u_{\sigma}(q)
\gamma^\mu
\, (\slashed q_2 + m)
\gamma^\nu
  u_{\sigma}(q)
+ \delta m^2_{\text{finite},2}
\ .
\eeq
The counterterm is represented as a Wick's diagram
in Fig.~\ref{fig:wickH}. The first term of
$\delta m^2_{t,X,2}$ is divergent as $t_r \to 0$
and cancels the divergent part of the integrand
in Eq.~(\ref{eq:dtm2t2}). The second term,
$\delta m^2_{\text{finite},2}$ is assumed independent
of $t_r$ and is added because a priori we do not know
what the finite part of the counterterm should be.
We discuss the choice of the finite part
in Sec.~\ref{sec:summary}.

\begin{figure}[h]
 \centering
 \includegraphics{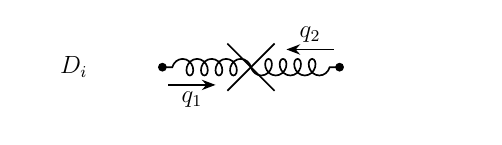}
 \caption{Wick's diagram for the gluon mass counterterm.}
 \label{fig:wickI}
\end{figure}

Next, we discuss the gluon self-energy terms that arise
from diagrams $D_f$ and $D_m$, Figs.~\ref{fig:wickF}
and \ref{fig:wickLMN}, Eqs.~(\ref{eq:ht2fsecond}) and
(\ref{eq:ht2m}), respectively. Both $\delta\tilde\mu_{t,q,2}^2$
and $\delta\tilde\mu_{t,g,2}^2$ diverge as $t_r \to 0$,
and the divergence can be traced to the $-1$ term in
the numerator of $B_{t,12(-3).(-1)(-2)3}^{q.q}$ and
$B_{t,12(-3).(-1)(-2)3}^{g.g}$, respectively.
Therefore, similarly to the quark case, we define
a counterterm,
\beq
\cH_{t,2}^{(i)}
\es
\int[q_3]
\theta(q_3^+)
\delta\tilde\mu_{t,X,2}^2(q_3)
\cN\left[ G^{i a}(-q_3) G^{i a}(q_3) \right] ,
\eeq
where
\beq
\delta\tilde\mu_{t,X,2}^2(q_3)
\es
2 T_f g^2
\int[q_1 q_2]
\frac{ \theta(q_1^+) \theta(q_2^+) }{ q_1^+ q_2^+ }
\, \tdelta_{12.3}
\frac{ (f_{t_r,12(-3)}^q)^2 }{ \left. q_1^- \right|_m + \left. q_2^- \right|_m - \left. q_3^- \right|_{m_g} }
\left[ \cM_{12}^2 - 2 (k^\perp)^2 \right]
\np
C_A
g^2
\int[q_1 q_2]
\frac{ \theta(q_1^+) \theta(q_2^+) }{ q_1^+ q_2^+ }
\, \tdelta_{12.3}
\frac{ (f_{t_r,12(-3)}^g)^2 }{ \left. q_1^- \right|_{m_g} + \left. q_2^- \right|_{m_g} - \left. q_3^- \right|_{m_g} }
2 (k^\perp)^2 \frac{ (1 - x_1 x_2)^2 }{ (x_1 x_2)^2 }
\np
\delta\mu_{\text{finite},2}^2
\ .
\label{eq:gluonmassX}
\eeq
The counterterm is represented as a Wick's diagram
in Fig.~\ref{fig:wickI}. The first term of
$\delta\tilde\mu_{t,X,2}^2(q_3)$ cancels the divergence
as $t_r \to 0$ of the quark loop self-energy,
Fig.~\ref{fig:wickF}, and the second term cancels
the divergence of the gluon loop self-energy,
Fig.~\ref{fig:wickLMN}. $\delta\mu_{\text{finite},2}^2$
is the finite part of the gluon mass counterterm,
discussed in Sec.~\ref{sec:summary}.

\begin{figure}
\centering
\includegraphics{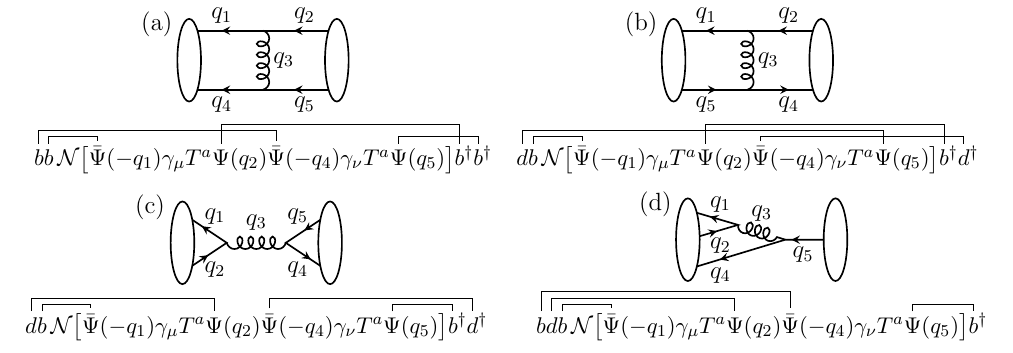}
\caption{\label{fig:matElA}Exemplary matrix elements
of $\cH_{t,2}^{(a)}$.}
\end{figure}

\begin{table}
\begin{tabular}{c|cccc}
$I$ & $a$ & $j$ & $k$ & $l$ \\
\hline
$K$ & $q$ & $q$ & $g$ & $g$ \\
$L$ & $q$ & $g$ & $q$ & $g$
\end{tabular}
\caption{\label{tab:IKL}For a given label (diagram) $I$,
labels $K$ and $L$ are determined from this table.}
\end{table}

Terms $\cH_{t,2}^{(I)}$ for $I \in \{a, j, k, l\}$,
corresponding to diagrams, Figs.~\ref{fig:wickA},
\ref{fig:wickJK}, and \ref{fig:wickLMN} have very
similar form, and can be written,
\beq
\cH_{t,2}^{(I)}
\es
\frac{S_I}{2}
\int[q_1 q_2 q_3 q_4 q_5]
\tdelta_{123} \tdelta_{45.3}
f^K_{t_r, 123} f^L_{t_r, 45(-3)}
\nn&&\ \times
\frac{ f_{t,1245}^{K.L} B_{t,123.45(-3)}^{K.L} }{ q_3^+ }
d_{\mu\nu}(q_3)
\cN\left[
\tilde J_{K}^{\mu\,c}(q_1, q_2)
\tilde J_{L}^{\nu\,c}(q_4, q_5)
\right] ,
\label{eq:cHIKLdjj}
\eeq
where $K$, and $L$ depend on $I$ in a way tabulated
in Table~\ref{tab:IKL}. In other words, for different $I$
only the labels $q$ and $g$ that label form factors,
RGPEP function $B$, and the currents are different.

The integrand can develop a singularity whenever
$q_3^+ = 0$. One of the sources of possible singularities
is the following expression:
\beq
\frac{ f_{t,1245}^{K.L} B_{t,123.45(-3)}^{K.L} }{ q_3^+ }
\es
  \left(
    f^K_{t, 123} f^L_{t, 45(-3)}
  - f_{t,1245}^{K.L}
  \right)
  \cF^{K.L}
\ ,
\label{eq:fBKL}
\eeq
where
\beq
\cF^{K.L}
\es
  \frac{1}{2}
  \left(
    \frac{1}{m_g^2 - (\left. q_1 \right|_K + \left. q_2 \right|_K )^2}
  + \frac{1}{m_g^2 - (\left. q_4 \right|_L + \left. q_5 \right|_L )^2}
  \right) ,
\eeq
and $\left. q \right|_K = \left. q \right|_{m_K}$ with
$m_K = m_q = m$ for $K = q$, and $m_K = m_g$ for $K = g$.
There are two cases, either $(q_1 + q_2)^2 \le 0$ or
$(q_1 + q_2)^2 \ge 4 m_K^2$. The former happens in
Figs.~\ref{fig:matElA}(a) and (b), while the latter
happens in Figs.~\ref{fig:matElA}(c) and (d).
Similarly, either $(q_4 + q_5)^2 \le 0$, see
Figs.~\ref{fig:matElA}(a), (b), and (d), or
$(q_4 + q_5)^2 \ge 4 m_L^2$, see Fig.~\ref{fig:matElA}(c).
We assume that $m_g$ is much smaller than the quark
mass and $4 m^2 > m_g^2$. Therefore, whether $m_K$
and $m_L$ are $m$ or $m_g$, $\cF^{K.L}$ and
Eq.~(\ref{eq:fBKL}) never develops any singularity
despite explicit $1/q_3^+$ factor.

Using momentum conservation and Eq.~(\ref{eq:qmuJmu}),
\beq
q_{3\mu} \tilde J_{K}^{\mu\,c}(q_1, q_2)
\es
  \frac{m_g^2 - (\left. q_1 \right|_K + \left. q_2 \right|_K)^2}{2 q_3^+}
  \tilde J_{K}^{+\,c}(q_1, q_2)
\ ,
\\
q_{3\nu} \tilde J_{L}^{\nu\,c}(q_4, q_5)
\es
  \frac{m_g^2 - (\left. q_4 \right|_L + \left. q_5 \right|_L)^2}{2 q_3^+}
  \tilde J_{L}^{+\,c}(q_4, q_5)
\ .
\eeq
Therefore, using Eq.~(\ref{eq:dmunu}),
\begin{multline}
d_{\mu\nu}(q_3)
\cN\left[
\tilde J_{K}^{\mu\,c}(q_1, q_2)
\tilde J_{L}^{\nu\,c}(q_4, q_5)
\right]
\rs
- g_{\mu\nu}
\cN\left[
\tilde J_{K}^{\mu\,c}(q_1, q_2)
\tilde J_{L}^{\nu\,c}(q_4, q_5)
\right]
\\
- \frac{(\left. q_1 \right|_K + \left. q_2 \right|_K)^2
+ (\left. q_4 \right|_L + \left. q_5 \right|_L)^2}{2 (q_3^+)^2}
\cN\left[
\tilde J_{K}^{+\,c}(q_1, q_2)
\tilde J_{L}^{+\,c}(q_4, q_5)
\right] .
\end{multline}
Next, we split $\cH_{t,2}^{(I)}$ into two terms:
\beq
\cH_{t,2}^{(I)}
\es
\cH_{t,2}^{(I,R)} + \cH_{t,2}^{(I,X)}
\ ,
\eeq
where $\cH_{t,2}^{(I,R)}$ is regular, while $\cH_{t,2}^{(I,X)}$
is singular:
\beq
\cH_{t,2}^{(I,R)}
\es
- \frac{S_I}{2}
\int[q_1 q_2 q_3 q_4 q_5]
\tdelta_{123} \tdelta_{45.3}
f^K_{t_r, 123} f^L_{t_r, 45(-3)}
\frac{ f_{t,1245}^{K.L} B_{t,123.45(-3)}^{K.L} }{ q_3^+ }
\nn&&\ \times
g_{\mu\nu}
\cN\left[
\tilde J_{K}^{\mu\,c}(q_1, q_2)
\tilde J_{L}^{\nu\,c}(q_4, q_5)
\right]
\nn&&
- \frac{S_I}{2}
\int[q_1 q_2 q_3 q_4 q_5]
\tdelta_{123} \tdelta_{45.3}
f^K_{t + t_r, 123} f^L_{t + t_r, 45(-3)}
\frac{ \tilde\cF^{K.L} }{ (q_3^+)^2 }
\nn&&\ \times
\cN\left[
\tilde J_{K}^{+\,c}(q_1, q_2)
\tilde J_{L}^{+\,c}(q_4, q_5)
\right]
 ,
\label{eq:Ht2IR}
\\
\cH_{t,2}^{(I,X)}
\es
\frac{S_I}{2}
\int[q_1 q_2 q_3 q_4 q_5]
\tdelta_{123} \tdelta_{45.3}
f^K_{t_r, 123} f^L_{t_r, 45(-3)}
f_{t,1245}^{K.L}
\frac{ \tilde\cF^{K.L} }{ (q_3^+)^2 }
\nn&&\ \times
\cN\left[
\tilde J_{K}^{+\,c}(q_1, q_2)
\tilde J_{L}^{+\,c}(q_4, q_5)
\right] .
\eeq
where
\beq
\tilde\cF^{K.L}
\es
\frac{(\left. q_1 \right|_K + \left. q_2 \right|_K)^2
+ (\left. q_4 \right|_L + \left. q_5 \right|_L)^2}{2}
\cF^{K.L}
\ .
\eeq
The first term of $\cH_{t,2}^{(I,R)}$ is not singular because
it does not contain any $1/q_3^+$ factors other than those
discussed in Eq.~(\ref{eq:fBKL}). The second term of
$\cH_{t,2}^{(I,R)}$ is not singular because it contains
the product of form factors, $f^K_{t, 123} f^L_{t, 45(-3)}$.
Note that
\beq
f^K_{t, 123} f^L_{t, 45(-3)}
\es
\exp\left(
-t \frac{[(\left. q_1 \right|_K + \left. q_2 \right|_K)^2 - m_g^2]^2}{(q_3^+)^2}
\right)
\exp\left(
-t \frac{[(\left. q_4 \right|_L + \left. q_5 \right|_L)^2 - m_g^2]^2}{(q_3^+)^2}
\right) .
\nn
\eeq
From previous analysis we know that
$[(q_1 + q_2)^2 - m_g^2]^2 \ge m_g^4$ and
$[(q_4 + q_5)^2 - m_g^2]^2 \ge m_g^4$, because
either $(q_1 + q_2)^2 \le 0$ or $(q_1 + q_2)^2
\ge 4 m_K^2 > m_g^2$, and the analogous holds
for $(q_4 + q_5)^2$. Therefore, as $q_3^+$ approaches
zero, $f^K_{t, 123} f^L_{t, 45(-3)}$ approaches zero
at least as quickly as $\exp[-t m_g^4(q_3^+)^{-2}]$
and can regulate any power of $1/q_3^+$. However,
this conclusion changes in the $m_g \to 0$ limit,
as discussed in Sec.~\ref{sec:casimir}.

In contrast, $\cH_{t,2}^{(I,X)}$ contains a singular
factor of $1/(q_3^+)^2$ which is regulated only by
$f^K_{t_r, 123} f^L_{t_r, 45(-3)}$. Therefore,
the matrix elements of $\cH_{t,2}^{(I,X)}$
diverge as $t_r \to 0$. Another singular term
in the effective Hamiltonian arises from the
instantaneous gluon term in the canonical
Hamiltonian. We define $\cH_{t,2}^{(0,JJ)} =
e^{-t(i\pd_f^-)^2} H_{JJ}$. For $I \in \{a, j, k, l\}$,
$\cH_{t,2}^{(0,JJ)} = \sum_{I} \cH_{t,2}^{(0,I)}$,
where
\beq
\cH_{t,2}^{(0,I)}
\es
\frac{S_I}{2}
\int[q_1 q_2 q_3 q_4 q_5]
\tdelta_{123} \tdelta_{45.3}
f^K_{t_r, 123} f^L_{t_r, 45(-3)}
\frac{ f_{t,1245}^{K.L} }{ (q_3^+)^2}
\cN\left[
\tilde J_{K}^{+\,c}(q_1, q_2)
\tilde J_{L}^{+\,c}(q_4, q_5)
\right] .
\nn
\eeq
Therefore,
\beq
\cH_{t,2}^{(0,I)} + \cH_{t,2}^{(I,X)}
\es
\frac{S_I}{2}
\int[q_1 q_2 q_3 q_4 q_5]
\tdelta_{123} \tdelta_{45.3}
f^K_{t_r, 123} f^L_{t_r, 45(-3)}
f_{t,1245}^{K.L}
\nn&&\times
\frac{ 1 + \tilde\cF^{K.L} }{ (q_3^+)^2 }
\cN\left[
\tilde J_{K}^{+\,c}(q_1, q_2)
\tilde J_{L}^{+\,c}(q_4, q_5)
\right] .
\eeq

We now determine the counterterm necessary to make the
matrix elements of these Hamiltonian terms finite
as $t_r \to 0$. Among diagrams in Fig.~\ref{fig:matElA},
only \ref{fig:matElA}(a) and \ref{fig:matElA}(b) need
to be studied. This is because $q_3^+ = 0$ implies
both $q_1^+ + q_2^+ = 0$ and $q_4^+ + q_5^+ = 0$.
If the sign of $q_1^+$ is the same as the sign of $q_2^+$,
as in diagrams with pair creation, then $q_1^+$ and $q_2^+$
need to vanish as $q_3^+ \to 0$. Similarly, if $q_4^+ q_5^+
< 0$, as in diagrams with pair annihilation, then both
$q_4^+ \to 0$ and $q_5^+ \to 0$ as $q_3^+ \to 0$. Since
wave functions are compactly supported, the matrix elements
cannot receive divergent contributions from diagrams with
pair creation or annihilation. In diagrams \ref{fig:matElA}(a)
and \ref{fig:matElA}(b) $q_1^+$ has an opposite sign
to the sign of $q_2^+$ and $q_4^+$ has an opposite sign
to the sign of $q_5^+$. Therefore, all four momenta
can remain nonzero as $q_3^+ \to 0$. Moreover, $q_1^-$,
$q_2^-$, $q_4^-$, and $q_5^-$ are all finite. Therefore,
\beq
(q_1 + q_2)^2
\es
(q_1^+ + q_2^+)(q_1^- + q_2^-) - (q_1^\perp + q_2^\perp)^2
\rs
- (q_3^\perp)^2
\ ,
\quad (q_3^+ = 0)
\\
(q_4 + q_5)^2
\es
(q_4^+ + q_5^+)(q_4^- + q_5^-) - (q_4^\perp + q_5^\perp)^2
\rs
- (q_3^\perp)^2
\ ,
\quad (q_3^+ = 0)
\eeq
which leads to
\beq
\tilde\cF^{K.L}
\es
- \frac{(q_3^\perp)^2}{m_g^2 + (q_3^\perp)^2}
\ .
\quad (q_3^+ = 0)
\label{eq:tFKLq30}
\eeq
The regulating factors are
\beq
f^K_{t_r, 123} f^L_{t_r, 45(-3)}
\es
\exp\left[
- 2 t_r \frac{[(q_3^\perp)^2 + m_g^2]^2}{(q_3^+)^2}
+ O\left(\frac{1}{q_3^+}\right)
\right] .
\quad (q_3^+ \to 0)
\label{eq:fKfLq30}
\eeq
Therefore, the singular factor $1/(q_3^+)^2$,
multiplied by $f^K_{t_r, 123} f^L_{t_r, 45(-3)}$,
gives in the leading order,
\beq
\int_{-\infty}^\infty
\frac{ dq_3^+ }{ (q_3^+)^2 }
e^{- 2 t_r \frac{[(q_3^\perp)^2 + m_g^2]^2}{(q_3^+)^2}}
\es
\frac{1}{m_g^2 + (q_3^\perp)^2}
\sqrt{\frac{\pi}{2 t_r}}
\ ,
\eeq
which is divergent in the $t_r \to 0$ limit. To counter
this divergence we define the counterterm,
\beq
\cH_{t,2}^{(0,I,X)}
\es
- \frac{S_I}{2}
\int[q_1 q_2 q_3 q_4 q_5]
\tdelta_{123} \tdelta_{45.3}
f_{t,1245}^{K.L}
\frac{m_g^2}{[m_g^2 + (q_3^\perp)^2]^2}
\sqrt{\frac{\pi}{2 t_r}}
\delta(q_3^+)
\nn&&\times
\,\cN\left[
\tilde J_{K}^{+\,c}(q_1, q_2)
\tilde J_{L}^{+\,c}(q_4, q_5)
\right] .
\label{eq:instGCT}
\eeq
To be more precise, the counterterm is $\cH_{0,2}^{(0,I,X)}$
and it is added to the initial condition at $t = 0$,
while $\cH_{t,2}^{(0,I,X)}$ is the term at $t > 0$
that is produced by the RGPEP flow in the second order
due to the presence of the counterterm. This counterterm
extends the one introduced in Ref.~\cite{Serafin:2023pkf}
to all Fock sectors and all combinations of quarks,
antiquarks, and gluons.

\begin{figure}
 \centering
 \includegraphics{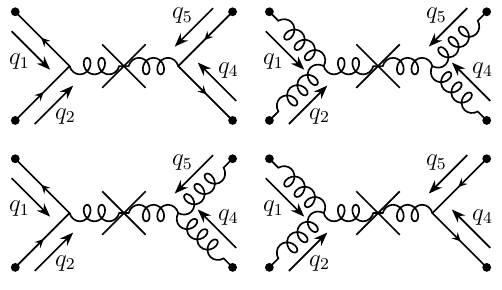}
 \caption{Wick's diagrams for the instantaneous
 interaction counterterms.}
 \label{fig:wickO}
\end{figure}

The instantaneous fermion diagrams are analyzed next.
The diagrams $D_b$ and $D_c$, Fig.~\ref{fig:wickBC} give,
\beq
\cH_{t,2}^{(b)} + \cH_{t,2}^{(c)}
\es
g^2
\int[q_1 q_2 q_3 q_5 q_6]
\, \tdelta_{123}
\, \tdelta_{56.2}
f_{t_r,123}^q
f_{t_r,(-2)56}^q
f_{t,1356}^{q.q}
\frac{ B_{t,123.(-2)56}^{q.q} }{ q_2^+ }
\nn&&\ \times
\cN\left[
\bar\Psi(-q_1) \gamma_\mu T^a G^{\mu a}(q_3)
\, (\slashed q_2 + m)
\gamma_\nu T^b G^{\nu b}(q_6) \Psi(q_5)
\right]
 .
\eeq
The diagram, Fig.~\ref{fig:instantaneous}(a) gives,
\beq
\cH_{t,2}^{(0,bc)}
\es
g^2
\int[q_1 q_2 q_3 q_5 q_6]
\, \tdelta_{123}
\, \tdelta_{56.2}
f_{t_r,123}^q
f_{t_r,(-2)56}^q
f_{t,1356}^{q.q}
\nn&&\ \times
\cN\left[
\bar\Psi(-q_1) \gamma_\mu T^a G^{\mu a}(q_3)
\, \frac{\gamma^+}{2 q_2^+}
\gamma_\nu T^b G^{\nu b}(q_6) \Psi(q_5)
\right]
 .
\eeq
The three diagrams together:
\beq
\cH_{t,2}^{(\Psi^2 G^2)}
\es
\cH_{t,2}^{(b)} + \cH_{t,2}^{(c)} + \cH_{t,2}^{(0,bc)}
\\
\es
g^2
\int[q_1 q_2 q_3 q_5 q_6]
\, \tdelta_{123}
\, \tdelta_{56.2}
f_{t_r,123}^q
f_{t_r,(-2)56}^q
f_{t,1356}^{q.q}
\nn&&\ \times
\cN\Bigg\{
\bar\Psi(-q_1) \gamma_\mu T^a G^{\mu a}(q_3)
\left[
  A \frac{\gamma^+}{2}
+ \frac{ B_{t,123.(-2)56}^{q.q} }{ q_2^+ }
  \left(
    \frac{1}{2} q_2^+ \gamma^-
  - q_2^\perp \gamma^\perp
  + m
  \right)
\right]
\nn&&\ \times
\gamma_\nu T^b G^{\nu b}(q_6) \Psi(q_5)
\Bigg\}
 ,
\label{eq:HPsi2G2first}
\eeq
where
\beq
A
\es
  \frac{1}{q_2^+}
+ \frac{ B_{t,123.(-2)56}^{q.q} }{ q_2^+ }
  \frac{m^2 + (q_2^\perp)^2}{q_2^+}
\label{eq:Afirst}
\\
\es
  \frac{1}{2}
  \left(
    \frac{- q_1^- - q_3^-}{(q_1 + q_3)^2 - m^2}
  + \frac{q_5^- + q_6^-}{(q_5 + q_6)^2 - m^2}
  \right)
\nn&&
- \left(
    \frac{1}{(q_1 + q_3)^2 - m^2}
  + \frac{1}{(q_5 + q_6)^2 - m^2}
  \right)
  \frac{f_{t,123}^q f_{t,(-2)56}^q}{f_{t,1356}^{q.q}}
  \frac{m^2 + (q_2^\perp)^2}{2 q_2^+}
\ .
\label{eq:Asecond}
\eeq
The only difference between this expression and the
expression Eq.~(127) in~\cite{Serafin:2025ouo} for Yukawa theory is that now we have gluons
instead of scalar fields. Gluonic fields introduce
some extra momentum dependence:
\beq
\gamma_\mu G^{\mu a}(q_3)
\es
  \frac{1}{2} \gamma^+ G^{- a}(q_3)
- \gamma^j G^{j a}(q_3)
\rs
  \left(
    \frac{q_3^j}{q_3^+} \gamma^+
  - \gamma^j
  \right)
  G^{j a}(q_3)
\ .
\eeq
This expression is singular only for $q_3^+ \to 0$, which
in this case corresponds to the external leg and is
regulated by compactness of the support of wave functions.
No new $q_2^+ = 0$ singularity is introduced. Hence,
the analysis proceeds as in~\cite{Serafin:2025ouo} for Yukawa theory. The expression
\beq
\frac{ B_{t,123.(-2)56}^{q.q} }{ q_2^+ }
\es
\frac{1}{2}
  \left(
    \frac{1}{(q_1 + q_3)^2 - m^2}
  + \frac{1}{(q_5 + q_6)^2 - m^2}
  \right)
  \left(
    1
  - \frac{f_{t,123}^q f_{t,(-2)56}^q}{f_{t,1356}^{q.q}}
  \right)
\label{eq:BqinstQ}
\eeq
does not contain singularities if the sign of $q_1^+$
is the same as the sign of $q_3^+$, i.e., $q_1^+ q_3^+ > 0$,
and the sign of $q_5^+$ is the same as the sign of $q_6^+$,
i.e., $q_5^+ q_6^+ > 0$, because $(q_1 + q_3)^2 \ge
(m + m_g)^2 > m^2$, and $(q_5 + q_6)^2 \ge (m + m_g)^2 > m^2$
in this case.

If $q_1^+ q_3^+ < 0$, then $(q_1 + q_3)^2 - m^2$ can become
zero, but at the same time $1 - f_{t,123}^q f_{t,(-2)56}^q
/ f_{t,1356}^{q.q}$ becomes zero as well and regulates
the expression, provided that $q_2^+ \neq 0$. However,
$q_2^+ = 0$ implies $(q_1 + q_3)^2 = - (q_2^\perp)^2$,
making $(q_1 + q_3)^2 - m^2$ nonzero. Therefore, $q_2^+$
has to be nonzero when $(q_1 + q_3)^2 - m^2$ is zero,
and the expression is indeed not singular. Analogous
argument applies to the $q_5^+ q_6^+ < 0$ case as well.

We have established that vanishing of $(q_1 + q_3)^2 - m^2$
or $(q_5 + q_6)^2 - m^2$ as in Eq.~(\ref{eq:BqinstQ})
does not lead to divergent matrix elements. The only
possible source of divergence left is when $q_2^+ \to 0$
in $A$ in the second line of Eq.~(\ref{eq:Asecond}).
This expression is also regular because
$f_{t,123}^q f_{t,(-2)56}^q$ has the form $\exp[
-(\text{positive number})/(q_2^+)^2]$, which makes the
expression go to zero as $q_2^+ \to 0$. Therefore,
no counterterm is needed for the fermion exchange
diagrams.

\section{Casimir operator}
\label{sec:casimir}

Having taken the $t_r\to 0$ limit in Sec.~\ref{sec:counterterms}
we now take the limit $m_g \to 0$. Our analysis involves
the quadratic Casimir operator of $SU(3)$, which we define
in terms of creation and annihilation operators. These
Casimir operators were previously used by one of us in the
analysis of states of tetraquarks~\cite{Kuang:2022vdy}.
The quadratic Casimir operator is
\beq
\hat C_2 =  \hat T^a \hat T^a
\ ,
\label{eq:defcasimir}
\eeq
where the sum over eight colors, $a = 1, \dots, 8$,
is implicit and
\beq
\hat T^a
\es
\sum_{\sigma, c_1, c_2}
\int\frac{dp^+ d^2p^\perp}{16\pi^3 p^+}
\theta(p^+)
[T_F^a]_{c_1 c_2}
b_{p \sigma c_1}^\dag b_{p \sigma c_2}
\np
\sum_{\sigma, c_1, c_2}
\int\frac{dp^+ d^2p^\perp}{16\pi^3 p^+}
\theta(p^+)
[T_{\bar F}^a]_{c_1 c_2}
d_{p \sigma c_1}^\dag d_{p \sigma c_2}
\np
\sum_{\sigma, c_1, c_2}
\int\frac{dp^+ d^2p^\perp}{16\pi^3 p^+}
\theta(p^+)
[T_A^a]_{c_1 c_2}
a_{p \sigma c_1}^\dag a_{p \sigma c_2}
\ ,
\label{eq:defTa}
\eeq
with
\beq
\left[ T_F^a \right]_{c_1 c_2}
\es
T^a_{c_1 c_2}
\ ,
\label{eq:defTF}
\\
\left[ T_{\bar F}^a \right]_{c_1 c_2}
\es
- T^a_{c_2 c_1}
\ ,
\label{eq:defTFbar}
\\
\left[ T_A^a \right]_{c_1 c_2}
\es
- i f^{a c_1 c_2}
\ .
\label{eq:defTA}
\eeq
Note that $c_1, c_2 \in \{1, 2, 3\}$ in Eqs.~(\ref{eq:defTF})
and (\ref{eq:defTFbar}), while $c_1, c_2 \in \{1, \dots, 8\}$
in Eq.~(\ref{eq:defTA}). Subscripts $F$, $\bar F$, and $A$
refer to $3$, $\bar 3$, and $8$ representations, respectively.
Representation $3$ is a fundamental representation associated
with quarks. Representation $\bar 3$ is a fundamental
representation associated with antiquarks and $\bar 3$ is
conjugate to $3$. Representation $8$ is the adjoint
representation and it is associated with gluons.
The color charge operators annihilate the vacuum
because their terms are normal ordered,
\beq
\hat T^a \ket{0} \es 0 \ .
\eeq
Moreover,
\beq
\left[ \hat T^a, b_{p \sigma c}^\dag \right]
\es
\sum_{c'}
\left[ T_F^a \right]_{c' c}
b_{p \sigma c'}^\dag
\ ,
\\
\left[ \hat T^a, d_{p \sigma c}^\dag \right]
\es
\sum_{c'}
\left[ T_{\bar F}^a \right]_{c' c}
d_{p \sigma c'}^\dag
\ ,
\\
\left[ \hat T^a, a_{p \sigma c}^\dag \right]
\es
\sum_{c'}
\left[ T_A^a \right]_{c' c}
a_{p \sigma c'}^\dag
\ .
\eeq
Therefore,
\beq
\hat C_2 b_{p \sigma c}^\dag \ket{0}
\es
C_F b_{p \sigma c}^\dag \ket{0} ,
\\
\hat C_2 d_{p \sigma c}^\dag \ket{0}
\es
C_F d_{p \sigma c}^\dag \ket{0} ,
\\
\hat C_2 a_{p \sigma c}^\dag \ket{0}
\es
C_A a_{p \sigma c}^\dag \ket{0} .
\eeq
where (keeping in mind implicit sum over $a$)
\beq
C_F
\es
\sum_{c' = 1}^3 \left[ T_F^a \right]_{c c'}
\left[ T_F^a \right]_{c' c}
\rs
\sum_{c' = 1}^3 \left[ T_{\bar F}^a \right]_{c c'}
\left[ T_{\bar F}^a \right]_{c' c}
\rs
\frac{N_c^2 - 1}{2 N_c} \ ,
\eeq
and
\beq
C_A
\es
\sum_{c' = 1}^8 \left[ T_A^a \right]_{c c'}
\left[ T_A^a \right]_{c' c}
\rs
N_c \ .
\eeq

Additionally, $g \hat T^a$ can be though of
as the charge operator. In terms of the color
current operators,
\beq
\hat T^a
\es
\frac{1}{2 g}
\left[
  J_{q}^{+\,a}(0)
+ 3 \tilde J_{g}^{+\,a}(0)
\right]
\\
\es
\frac{1}{2g} \int[q]
\left[ \tilde J_q^{+ \, a}(-q, q)
+ 3 \tilde J_g^{+ \, a}(-q, q) \right]
\label{eq:TaEqtJtJ}
\\
\es
\frac{1}{2}
\int[q]
\,\cN\left[
\bar\Psi(q) \gamma^+ T^a \Psi(q)
\right]
+
\frac{1}{2}
\int[q]
(-i f^{c_1c_2a})
q^+
G^{j c_1}(-q) G^{j c_2}(q)
\ ,
\label{eq:TaJ}
\eeq
where we assume $\epsilon^+ = 0$.

In accordance with Wick's theorem, the Casimir operator
splits into three normal-ordered terms,
\beq
\hat C_2
\es
\cN\left(\hat T^a \hat T^a\right)
+ 
\int[q]
\frac{C_F |q^+|}{q^+}
\,\cN\left[ \bar\Psi(q) \frac{\gamma^+}{2} \Psi(q) \right]
+
\int[q]
\theta(q^+)
C_A q^+
G^{j a}(-q)
G^{j a}(q)
\ .
\nn
\label{eq:C2wick}
\eeq
The first term corresponds to gluon exchange interactions
between color currents, while the other two terms correspond
to mass terms in the Hamiltonian. Comparison with the free
part of the Hamiltonian, Eq.~(\ref{eq:freeCan}), reveals
that $C_F |q^+|$ and $C_A |q^+|$ have the correct form for
a correction to the quark mass $m^2$, and gluon mass $m_g^2$,
respectively. Therefore these are the factors that we expect
to arise from the self-interaction terms. We first consider
gluon self-interactions, and then quark self-interactions,
and finally interactions between color currents.

Gluon self interactions have three contributions:
unrenormalized gluon loop, Eq.~(\ref{eq:delTildMuG}),
unrenormalized quark loop, Eq.~(\ref{eq:delTildMuQ}),
and the gluon mass counterterm, Eq.~(\ref{eq:gluonmassX}).
We define the renormalized gluon loop contribution,
$\delta\mu_{t,g,2}^2(q_3)$ as the sum of
$\delta\tilde\mu_{t,g,2}^2(q_3)$ and the second term
in Eq.~(\ref{eq:gluonmassX}):
\beq
\delta\mu_{t,g,2}^2(q_3)
\es
C_A g^2
\int[q_1 q_2]
\frac{ \theta(q_1^+) }{ |q_1^+| }
\frac{ \theta(q_2^+) }{ |q_2^+| }
\, \tdelta_{12.3}
q_3^+ \frac{ e^{-2t\frac{\cM_{12}^2 - m_g^2}{(q_3^+)^2}} }{ \cM_{12}^2 - m_g^2 }
\cdot
2 k^2 \frac{ (1 - x_1 x_2)^2 }{ (x_1 x_2)^2 }
\ ,
\eeq
where $x_1$, $x_2$, and $k$ are defined according to
Eqs.~(\ref{eq:fermRel1}) and (\ref{eq:fermRel2}), and
\beq
\cM_{12}^2
\es
\frac{m_g^2 + k^2}{x_1 x_2}
\ .
\eeq
We also define the renormalized quark loop contribution,
$\delta\mu_{t,q,2}^2(q_3)$ as the sum of
$\delta\tilde\mu_{t,q,2}^2(q_3)$ and
the first term in Eq.~(\ref{eq:gluonmassX}):
\beq
\delta\mu_{t,q,2}^2(q_3)
\es
2 T_f g^2
\int[q_1 q_2]
\frac{ \theta(q_1^+) }{ |q_1^+| }
\frac{ \theta(q_2^+) }{ |q_2^+| }
\, \tdelta_{12.3}
q_3^+ \frac{ e^{-2t\frac{\cM_{12}^2 - m_g^2}{(q_3^+)^2}} }{ \cM_{12}^2 - m_g^2 }
\left( \cM_{12}^2 - 2 k^2 \right) ,
\eeq
where
\beq
\cM_{12}^2
\es
\frac{m^2 + k^2}{x_1 x_2}
\ .
\eeq
$\delta\mu_{t,q,2}^2(q_3)$ is finite in the $m_g \to 0$
limit. For $m^4 t \ll (q_3^+)^2$ it takes particularly
simple form,
\beq
\lim_{m_g \to 0}
\delta\mu_{t,q,2}^2(q_3)
\es
\frac{g^2}{16\pi^2}
|q_3^+|
\sqrt{\frac{\pi}{2t}}
\cdot\frac{1}{3}
+ o(1)
\ .
\quad
\left( m \to 0 \right)
\eeq
The gluon loop contribution on the other hand
is logarithmically divergent,
\beq
\delta\mu_{t,g,2}^2(q_3)
\es
C_A |q_3^+|
\frac{g^2}{16\pi^2}
\sqrt{\frac{\pi}{2t}}
\left[
  \log{\left(\frac{(q_3^+)^2}{8 m_g^4 t}\right)}
- \gamma
- \frac{23}{6}
\right]
+ o(1)
\ .
\quad
(m_g \to 0)
\label{eq:delmu2tg2}
\eeq
Renormalized quark self interaction due to quark-gluon
loop is logarithmically divergent as well,
\beq
\delta m^2_{t,2}
\es
\delta\tilde m^2_{t,2} + \delta m^2_{t,X,2}
\rs
C_F |q_1^+|
\frac{g^2}{16\pi^2} \sqrt{\frac{\pi}{2t}}
\log{\left(\frac{1}{16 m_g^4} \right)}
+ O(1)
\ .
\quad
(m_g \to 0)
\eeq
For $m^4 t \ll (q_1^+)^2$,
the expression can be evaluated further,
\beq
\delta m^2_{t,2}
\es
C_F |q_1^+|
\frac{g^2}{16\pi^2} \sqrt{\frac{\pi}{2t}}
\left[
\log{\left(\frac{(q_1^+)^2}{8 m_g^4 t}\right)}
- \gamma
- \frac{7}{2}
\right]
+ o(1)
\ .
\quad
(m_g \to 0, m \to 0)
\label{eq:delmu2tq2}
\eeq

The gluon self-energy and quark self-energy are composed of
the expected color-momentum factors, $C_A |q_3^+|$ and
$C_F |q_1^+|$, respectively, times a common divergent
factor $\frac{g^2}{16\pi^2} \sqrt{\frac{\pi}{2t}}
\log{\left(\frac{1}{m_g^4} \right)}$. The same
common factor is also found in the gluon exchange
terms. To show it we focus on the second term
of $\cH_{t,2}^{(I,R)}$, which contains
$\tilde J_{K}^{+\,c} \tilde J_{L}^{+\,c}$,
see Eq.~(\ref{eq:Ht2IR}). The only potential
source of divergence is the vicinity of $q_3^+ = 0$.
Using Eq.~(\ref{eq:tFKLq30}), we replace
$\tilde\cF^{K.L} \to -1$. The $m_g$ term in
the denominator of Eq.~(\ref{eq:tFKLq30}) can
be neglected, because it leads to small
corrections as $m_g \to 0$. Using Eq.~(\ref{eq:fKfLq30}),
$f^K_{t, 123} f^L_{t, 45(-3)}$ is replaced with
its approximation valid at $q_3^+ = 0$. Nonzero $m_g$
needs to be kept here, because it regulates the
logarithmic divergence. After these simplifications
we arrive at
\beq
J^{(I)}
\es
\frac{S_I}{2}
\int[q_1 q_2 q_3 q_4 q_5]
\theta(-q_1^+ q_2^+)
\theta(-q_4^+ q_5^+)
\tdelta_{123}
\tdelta_{45.3}
\exp\left(
- 2 t \frac{[m_g^2 + (q_3^\perp)^2]^2}{(q_3^+)^2}
\right)
\nn&&\ \times
\frac{1}{(q_3^+)^2}
\cN\left[
\tilde J_{K}^{+\,c}(q_1, q_2)
\tilde J_{L}^{+\,c}(q_4, q_5)
\right]
 ,
\eeq
where one more modification to the discussed term
of $\cH_{t,2}^{(I,R)}$ has been made: we inserted
$\theta(-q_1^+ q_2^+)$, which ensures that $q_1^+$
and $q_2^+$ have different signs, and
$\theta(-q_4^+ q_5^+)$, which ensures that $q_4^+$
and $q_5^+$ have different signs. This is necessary
to exclude pair creation and pair annihilation
interaction terms, which are regulated by the
compactness of the of the support of wave functions.
Therefore, $J^{(I)}$ describes only gluon exchange
between two color currents.

To evaluate $J^{(I)}$ further we change integration
variables from $q_1$, $q_2$, $q_4$, and $q_5$ to
$P_{12}$, $Q_{12}$, $P_{45}$, and $Q_{45}$, where
\beq
P_{12}^{+,\perp}
\es
\frac{q_2^{+,\perp} - q_1^{+,\perp}}{2}
\ ,
\\
Q_{12}^{+,\perp}
\es
q_1^{+,\perp} + q_2^{+,\perp}
\ ,
\\
P_{45}^{+,\perp}
\es
\frac{q_5^{+,\perp} - q_4^{+,\perp}}{2}
\ ,
\\
Q_{45}^{+,\perp}
\es
q_4^{+,\perp} + q_5^{+,\perp}
\ .
\eeq
We can easily evaluate integration over $Q_{45}$,
and $Q_{12}$, because momentum conservation forces
$Q_{45} = -Q_{12} = q_3$. The two theta functions
become $\theta[(P_{12}^+)^2 - (q_3^+)^2/4]$,
and $\theta[(P_{45}^+)^2 - (q_3^+)^2/4]$ and can
be replaced with one that produces the same result:
$\theta(P^+ - |q_3^+|)$, where $P^+ =
2\min(|P_{12}^+|, |P_{45}^+|) =
\min(|q_2^+ - q_1^+|, |q_5^+ - q_4^+|)$.
Therefore,
\beq
J^{(I)}
\es
\frac{S_I}{2}
\int[P_{12} P_{45}]
\int[q_3]
\theta\left( P^+ - |q_3^+| \right)
\exp\left(
- 2 t \frac{[m_g^2 + (q_3^\perp)^2]^2}{(q_3^+)^2}
\right)
\frac{1}{(q_3^+)^2}
\nn&&\ \times
\cN\left[
\tilde J_{K}^{+\,c}\left(
- \frac{q_3}{2} - P_{12},
- \frac{q_3}{2} + P_{12}
\right)
\tilde J_{L}^{+\,c}\left(
  \frac{q_3}{2} - P_{45},
  \frac{q_3}{2} + P_{45}
\right)
\right] .
\eeq
The divergence is due to the behavior of the integrand
at $q_3^+ = 0$, $q_3^\perp = 0$, as $m_g \to 0$.
To extract it we Taylor expand $\tilde J_{K}^{+\,c}
\tilde J_{L}^{+\,c}$ in $q_3$. Only the first term,
\beq
J^{(I)}_\text{div}
\es
\frac{S_I}{2}
\int[P_{12} P_{45}]
\int[q_3]
\theta\left( P^+ - |q_3^+| \right)
\exp\left(
- 2 t \frac{[(q_3^\perp)^2 + m_g^2]^2}{(q_3^+)^2}
\right)
\frac{1}{(q_3^+)^2}
\nn&&\ \times
\cN\left[
\tilde J_{K}^{+\,c}\left( - P_{12}, P_{12} \right)
\tilde J_{L}^{+\,c}\left( - P_{45}, P_{45} \right)
\right]
\label{eq:divFirstEq}
\eeq
is divergent. Integration over $q_3$ gives
\beq
&&
\int[q_3]
\theta\left( P^+ - |q_3^+| \right)
\exp\left(
- 2 t \frac{[(q_3^\perp)^2 + m_g^2]^2}{(q_3^+)^2}
\right)
\frac{1}{(q_3^+)^2}
\nn
\es
\frac{1}{2}
\frac{1}{16\pi^2}
\sqrt{\frac{\pi}{2t}}
\left[
\log\left(\frac{(P^+)^2}{8 m_g^4 t}\right)
- \gamma
\right]
+ O\left(m_g^2\right) .
\eeq
Note that the above result depends on $P_{12}^+$
and $P_{45}^+$ through $P^+$. To obtain $\hat T^a$
in Eq.~(\ref{eq:divFirstEq}) we need to work around
this dependence, because Eq.~(\ref{eq:TaEqtJtJ})
does not contain any $\log|q^+|$ factors.
Fortunately we can replace $P^+$ with some $\cP^+$
that does not depend on neither $P_{12}^+$ or $P_{45}^+$,
because $\log[(P^+)/m_g^4] = \log[(\cP^+)/m_g^4]
+ \log(P^+/\cP^+)$ and $\log(P^+/\cP^+)$ is $O(1)$
as $m_g \to 0$. Therefore,
\beq
J^{(I)}_\text{div}
\es
\frac{g^2}{16\pi^2}
\sqrt{\frac{\pi}{2t}}
\log\left(\frac{(\cP^+)^2}{8 m_g^4 t}\right)
\cN\left(
\hat T_{K}^c
\hat T_{L}^c
\right)
+ O\left(1\right) ,
\quad(m_g \to 0)
\eeq
where
\beq
\hat T_q^a
\es
\frac{1}{2g} \int[q] \tilde J_q^{+ \, a}(-q, q)
\ ,
\\
\hat T_g^a
\es
\frac{3}{2g} \int[q] \tilde J_g^{+ \, a}(-q, q)
\ .
\eeq
Note that the appropriate factors of 3 for
$\hat T_g^a$ come from the symmetry factor, $S_I$.
Summing $\hat T_{K}^c \hat T_{L}^c$ over all four
values of $I \in \{ a, j, k, l \}$ gives
$\hat T^c \hat T^c$.

To recover all three terms of Eq.~(\ref{eq:C2wick})
with a common diverging factor we replace logarithms
of momenta in Eqs.~(\ref{eq:delmu2tg2}) and
(\ref{eq:delmu2tq2}) with $\log\cP^+$. Therefore,
the effective Hamiltonian is,
\beq
\cH_{t,2}
\es
\frac{g^2}{16\pi^2}
\sqrt{\frac{\pi}{2t}}
\left[\log\left(\frac{(\cP^+)^2}{8 m_g^4 t}\right)
- \gamma \right]
\hat T^c
\hat T^c
+ O\left(1\right) .
\quad(m_g \to 0)
\label{eq:effcasimir}
\eeq
We see from Eq.~(\ref{eq:effcasimir}) that the logarithmically
divergent term in the effective Hamiltonian as $m_g \to 0$ is
proportional to the Casimir operator, Eq.~(\ref{eq:defcasimir}).
As a consequence the divergent term vanishes in the color
singlet sector, which is the eigenvalue zero eigenspace
of the Casimir operator. This result generalizes
the cancellation of infrared divergences observed
in the quark-antiquark sector in Ref.~\cite{Serafin:2023pkf}.
The same kind of cancellation has been pointed out
in Ref.~\cite{Perry:1994mv}, but to our knowledge
no Casimir operator has been previously reported.

It is worth pointing out that even though matrix elements
of $\cH_{t,2}$ are logarithmically divergent between
states that are not color singlets, color nonsinglet
quantities that are produced from these matrix elements
might still be well-defined. For example, single-quark mass
eigenvalue is well-defined in the $m_g \to 0$ limit as long
as it is computed using bound state perturbation theory up
to second order with $m$ taken as the unperturbed mass.
To obtain the mass eigenvalue one has to add the second-order
mass term from $\cH_{t,2}$ and the product of two first-order
vertices, as per bound state perturbation theory.
The summation leads to the cancellation of $m_g$-divergent
terms~\cite{Serafin:2023pkf}. Nonperturbatively, this
cancellation may be incomplete, which would lead
to divergent quark eigenmass. Such lack of cancellation
would simply mean that the perturbation theory with $m$
as the unperturbed eigenmass is not the proper setup.
In Ref.~\cite{Serafin:2023pkf} the obstruction to
the cancellation is provided by the gluon mass ansatz.

The same gluon exchange Hamiltonian term that contributes
to the Casimir operator also provides a logarithmic
confining potential at large
separations~\cite{Perry:1994mv,Serafin:2023pkf}.
It is very satisfying that confinement removes the divergences
associated with vanishing gluon mass. The alternative would
be the introduction of counter terms arising from finite
gluon mass. One would expect that the finite part of such
counter terms may cause the effective Hamiltonian to deviate
from QCD, and so the absence of such counter terms is a positive
feature of this effective Hamiltonian. We have only shown
that this is true to second order in the solution of the RGPEP
equations. It is an interesting question whether this is a general
feature of gluon mass regularization on the light-front,
i.e. whether the divergent mass terms are proportional
to Casimir operators, or functions thereof, at all
orders in RGPEP.

\section{Summary of the effective Hamiltonian}
\label{sec:summary}

The mass terms of the effective Hamiltonian are:
\beq
\cH_\text{mass}
\es
  \int[q]
  \frac{m_t^2(q) + (q^\perp)^2}{q^+}
  \cN\left[
  \bar\Psi(q)
  \frac{\gamma^+}{2}
  \Psi(q)
  \right]
\np
  \int[q]
  \theta(q^+)
  \frac{\mu_t^2(q) + (q^\perp)^2}{q^+}
  \cN\left[ G^{i\,a}(-q) G^{i\,a}(q) \right] ,
\eeq
where
\beq
m^2_t(q)
\es
m^2
+
C_F g^2
\int[q_2 q_3]
\, \tdelta_{23.q}
\frac{ \theta(q_2^+) \theta(q_3^+) }{ q_2^+ q_3^+ }
\frac{ (f_{t,(-q)23}^q)^2 }
{ \left. q_2^- \right|_{m} + \left. q_3^- \right|_{m_g}
- \left. q_1^-\right|_{m} }
\nt
d_{\mu\nu}(q_3)
  \bar u_{\sigma}(q)
\gamma^\mu
\, (\slashed q_2 + m)
\gamma^\nu
  u_{\sigma}(q)
+ \delta m^2_{\text{finite},2}
\ .
\eeq
\beq
\mu_t^2(q)
\es
m_g^2
+
2 T_f g^2
\int[q_1 q_2]
\frac{ \theta(q_1^+) \theta(q_2^+) }{ q_1^+ q_2^+ }
\, \tdelta_{12.q}
\frac{ (f_{t,12(-q)}^q)^2 }{ \left. q_1^- \right|_m + \left. q_2^- \right|_m - \left. q^- \right|_{m_g} }
\left[ \cM_{12}^2 - 2 (k^\perp)^2 \right]
\np
C_A
g^2
\int[q_1 q_2]
\frac{ \theta(q_1^+) \theta(q_2^+) }{ q_1^+ q_2^+ }
\, \tdelta_{12.q}
\frac{ (f_{t,12(-q)}^g)^2 }{ \left. q_1^- \right|_{m_g} + \left. q_2^- \right|_{m_g} - \left. q^- \right|_{m_g} }
2 (k^\perp)^2 \frac{ (1 - x_1 x_2)^2 }{ (x_1 x_2)^2 }
\np
\delta\mu_{\text{finite},2}^2
\ .
\eeq
In our previous analysis of RGPEP for the Yukawa
model~\cite{Serafin:2025ouo} we fixed the analogous
finite part of the fermion counterterm by demanding
the physical mass of the fermion to be equal to the
unperturbed mass in the Hamiltonian. In the second
order of perturbation theory the physical mass
receives contributions from the second-order
self-energy terms in the Hamiltonian, such as those
in Eq.~(\ref{eq:ht2de}), and from a product of two
first-order vertices arising from bound state
perturbation theory. Our condition forces these
two contributions to cancel exactly. If we were
to use the same condition here, then
$\delta m^2_{\text{finite},2}
= \delta\mu_{\text{finite},2}^2 = 0$. Strictly speaking,
we cannot use the same argument because a single
free quark and a single free gluon are forbidden
by color confinement. However, this is the unique
choice which ensures $m^2_t(q) \to m^2$, and
$\mu_t^2(q) \to m_g^2$, as $t \to \infty$, which
is consistent with coupling coherence, see discussion
below Eq.~(143) in Ref.~\cite{Perry:2001je}.
For the time being we leave the finite parts
of the mass counterterms unspecified.

The first order vertices give rise to terms:
\beq
\cH_{t,1}
\es
g
\int[q_1 q_2 q_3]
\,\tdelta_{123}
\, f^q_{t,123}
\,\cN\left[
\bar\Psi(-q_1) \gamma_\mu T^a \Psi(q_2) G^{\mu a}(q_3)
\right]
\np
\frac{g}{3!}
\int[q_1 q_2 q_3]
\,\tdelta_{123}
\, f^g_{t,123}
F_{\alpha\beta\gamma}^{abc}(q_1, q_2, q_3)
\, G^{\alpha a}(q_1) G^{\beta b}(q_2) G^{\gamma c}(q_3)
\ .
\eeq

The second order vertices give rise to terms:
\beq
\cH_{t,2}^{(G^4)}
\es
\frac{g^2}{4!}
\int[q_1 q_2 q_3 q_4]
\,f_{t,1234}
\,\tdelta_{1234}
\,F_{\alpha\beta\gamma\delta}^{abcd}
\,\cN\left[
G^{\alpha\,a}(q_1)
G^{\beta\,b}(q_2)
G^{\gamma\,c}(q_3)
G^{\delta\,d}(q_4)
\right] ,
\eeq

The Hamiltonian includes current-current interactions:
\beq
\cH_{t,2}^{(KL)}
\es
  \cH_{t,2}^{(I,R)}
+ \cH_{t,2}^{(0,I)}
+ \cH_{t,2}^{(I,X)}
+ \cH_{t,2}^{(0,I,X)}
\ ,
\eeq
where
\beq
\cH_{t,2}^{(I,R)}
\es
- \frac{S_I}{2}
\int[q_1 q_2 q_3 q_4 q_5]
\tdelta_{123} \tdelta_{45.3}
\frac{ f_{t,1245}^{K.L} B_{t,123.45(-3)}^{K.L} }{ q_3^+ }
g_{\mu\nu}
\cN\left[
\tilde J_{K}^{\mu\,c}(q_1, q_2)
\tilde J_{L}^{\nu\,c}(q_4, q_5)
\right]
\nn&&
- \frac{S_I}{2}
\int[q_1 q_2 q_3 q_4 q_5]
\tdelta_{123} \tdelta_{45.3}
f^K_{t, 123} f^L_{t, 45(-3)}
\frac{ \tilde\cF^{K.L} }{ (q_3^+)^2 }
\cN\left[
\tilde J_{K}^{+\,c}(q_1, q_2)
\tilde J_{L}^{+\,c}(q_4, q_5)
\right] ,
\eeq
and
\beq
\cH_{t,2}^{(0,I)} + \cH_{t,2}^{(I,X)}
\es
\frac{S_I}{2}
\int[q_1 q_2 q_3 q_4 q_5]
\tdelta_{123} \tdelta_{45.3}
f^K_{t_r, 123} f^L_{t_r, 45(-3)}
f_{t,1245}^{K.L}
\nn&&\times
\frac{ 1 + \tilde\cF^{K.L} }{ (q_3^+)^2 }
\cN\left[
\tilde J_{K}^{+\,c}(q_1, q_2)
\tilde J_{L}^{+\,c}(q_4, q_5)
\right] ,
\\
\cH_{t,2}^{(0,I,X)}
\es
- \frac{S_I}{2}
\int[q_1 q_2 q_3 q_4 q_5]
\tdelta_{123} \tdelta_{45.3}
f_{t,1245}^{K.L}
\frac{m_g^2}{[m_g^2 + (q_3^\perp)^2]^2}
\sqrt{\frac{\pi}{2 t_r}}
\delta(q_3^+)
\nn&&\times
\,\cN\left[
\tilde J_{K}^{+\,c}(q_1, q_2)
\tilde J_{L}^{+\,c}(q_4, q_5)
\right] .
\eeq

The final term in the effective Hamiltonian is a seagull term of the following form:
\beq
\cH_{t,2}^{(\Psi^2 G^2)}
\es
g^2
\int[q_1 q_2 q_3 q_5 q_6]
\, \tdelta_{123}
\, \tdelta_{56.2}
f_{t,1356}^{q.q}
\nt
\cN\Bigg\{
\bar\Psi(-q_1) \gamma_\mu T^a G^{\mu a}(q_3)
\left[
  \frac{\gamma^+}{2 q_2^+}
+ \frac{ B_{t,123.(-2)56}^{q.q} }{ q_2^+ }
  ( \slashed q_2 + m )
\right]
\gamma_\nu T^b G^{\nu b}(q_6) \Psi(q_5)
\Bigg\} .
\nn
\eeq
This second order RGPEP Hamiltonian for light-front QCD is rather compact. In section~\ref{sec:conclusion} we will discuss some of the properties of this Hamiltonian. 

\section{Conclusion}
\label{sec:conclusion}

In this work we have derived the second order RGPEP Hamiltonian for light-front QCD. The resulting Hamiltonian has many interesting features. Once ultraviolet renormalization is performed no new infrared divergences due to the finite gluon mass as a regulator are found. This is because the divergent terms involving the gluon mass are proportional to the quadratic Casimir operator of $SU(3)$ and hence vanish in the color singlet sector of physical states. The Hamiltonian includes a logarithmic confining potential, which has previously reported by Perry in~\cite{Perry:1994mv} and by Serafin~\cite{Serafin:2023pkf}. The Hamiltonian is relatively compact, which is promising from the point of view future numerical work and of resource estimates for quantum computation of the kind previously carried out for the RGPEP Yukawa Hamiltonian in~\cite{gustin2026renormalized}.  

Naturally the physical validity of this RGPEP Hamiltonian can only be established through further analytic and computational work.  A natural theoretical question is whether $\cH_t$ lead to perturbation theory that is the same as the standard perturbative QCD? Whether the RGPEP Hamiltonian at second order is sufficient to capture the full physics of QCD, including confinement and chiral symmetry breaking remains to be seen. 

The development of this Hamiltonian made further use of the Wick diagram techniques introduced previously in~\cite{Serafin:2025ouo}. We hope that these techniques will enable RGPEP calculations to higher order if we find that the second order Hamiltonian does not contain all the physics of QCD. Finally, we hope that this and future work will shed further light on questions of renormalizability on the light-front.

\bibliographystyle{apsrev4-1}
\bibliography{citations}

\end{document}